\documentclass[12pt]{article}
\usepackage{amsmath,amssymb,amsthm,amsxtra,overpic,bbm,bm,epsfig,ulem}
\usepackage{color}
\usepackage{cite}

\usepackage{hyperref}
\usepackage{url}

\def\thefootnote{\fnsymbol{footnote}}

\begin{document}

\vspace{0.2cm}

\begin{center}
{\large\bf The translational $\mu$-$\tau$ reflection symmetry of Majorana neutrinos}
\end{center}

\vspace{0.2cm}

\begin{center}
{\bf Zhi-zhong Xing$^{1,2,3}$}
\footnote{E-mail: xingzz@ihep.ac.cn}
\\
{\small $^{1}$Institute of High Energy Physics, Chinese Academy of Sciences,
Beijing 100049, China \\
$^{2}$School of Physical Sciences, University of Chinese Academy of Sciences,
Beijing 100049, China \\
$^{3}$Center of High Energy Physics, Peking University, Beijing 100871, China}
\end{center}

\vspace{2cm}
\begin{abstract}
The present neutrino oscillation data allow $m^{}_1 = 0$ (or
$m^{}_3 = 0$) for the neutrino mass spectrum and support $\theta^{}_{23} \simeq \pi/4$
and $\delta \simeq -\pi/2$ as two good approximations for the
PMNS lepton flavor mixing matrix $U$. We show that these intriguing possibilities can
be a very natural consequence of the {\it translational} $\mu$-$\tau$ reflection
symmetry --- the effective Majorana neutrino mass term keeps invariant under the
transformations $\nu^{}_{e \rm L} \to (\nu^{}_{e \rm L})^c + U^*_{e i} z^{c}_\nu$,
$\nu^{}_{\mu \rm L} \to (\nu^{}_{\tau \rm L})^c + U^*_{\tau i} z^{c}_\nu$ and
$\nu^{}_{\tau \rm L} \to (\nu^{}_{\mu \rm L})^c + U^*_{\mu i} z^{c}_\nu$ (for
$i = 1$ or $3$), where $z^{c}_\nu$ is the charge conjugation of a constant spinor
field $z^{}_\nu$. Extending such a working flavor symmetry to the canonical seesaw
mechanism at a superhigh-energy scale, we calculate its soft breaking effects
at the electroweak scale by using the one-loop renormalization-group equations. \\

\end{abstract}

\newpage

\def\thefootnote{\arabic{footnote}}
\setcounter{footnote}{0}

\section{Introduction}

The facts that the three known (active) neutrinos have nonzero but tiny masses and
lepton flavors are significantly mixed~\cite{Zyla:2020zbs} tell us that the standard
model (SM) of particle physics is incomplete at least in its flavor aspects, although
this great theory of electromagnetic and weak interactions has proved to be a huge
success in describing how the Universe works. A particular characteristic of
massive neutrinos is the flavor oscillation --- a pure quantum phenomenon,
as their flavor eigenstates $\nu^{}_\alpha$ (for $\alpha = e, \mu, \tau$) are
the quantum superpositions of their mass eigenstates $\nu^{}_i$ (for $i = 1, 2, 3$):
\begin{eqnarray}
\left(\begin{matrix} \nu^{}_e \cr \nu^{}_\mu \cr \nu^{}_\tau \cr \end{matrix}
\right)_{\hspace{-0.07cm} \rm L}
= \left(\begin{matrix} U^{}_{e 1} & U^{}_{e 2} & U^{}_{e 3} \cr
U^{}_{\mu 1} & U^{}_{\mu 2} & U^{}_{\mu 3} \cr
U^{}_{\tau 1} & U^{}_{\tau 2} & U^{}_{\tau 3} \cr \end{matrix} \right)
\left(\begin{matrix} \nu^{}_1 \cr \nu^{}_2 \cr \nu^{}_3 \cr \end{matrix}
\right)_{\hspace{-0.07cm} \rm L} \; ,
\label{eq:1}
%     (1)
\end{eqnarray}
where $U^{}_{\alpha i}$ stand for the Pontecorvo-Maki-Nakagawa-Sakata (PMNS) lepton
flavor mixing matrix elements~\cite{Pontecorvo:1957cp,Maki:1962mu,Pontecorvo:1967fh}.
A global analysis of current neutrino oscillation data in the three-flavor
scheme leaves us with the following information on neutrino masses, lepton
flavor mixing and CP violation~\cite{Capozzi:2021fjo,Gonzalez-Garcia:2021dve}:
\begin{itemize}
\item     The mass ordering of three neutrinos can be either normal (namely,
$m^{}_1 < m^{}_2 < m^{}_3$) or inverted (namely, $m^{}_3 < m^{}_1 < m^{}_2$),
but the latter is slightly disfavored as compared with the former. Now that
the absolute neutrino mass scale has been undetermined, the possibility of
$m^{}_1 = 0$ or $m^{}_3 = 0$ is actually allowed as an extreme but intriguing
case of the neutrino mass spectrum.

\item     The $3\times 3$ PMNS matrix $U$ exhibits an
approximate but suggestive $\mu$-$\tau$ interchange symmetry $|U^{}_{\mu i}|
\simeq |U^{}_{\tau i}|$ (for $i = 1, 2, 3$), which is rather different from
the observed pattern of the Cabibbo-Kobayashi-Maskawa (CKM) quark flavor mixing
matrix~\cite{Cabibbo:1963yz,Kobayashi:1973fv}. This observation, together with
the preliminary experimental evidence for significant CP violation in neutrino
oscillations~\cite{T2K:2019bcf,NOvA:2019cyt}, implies that $\theta^{}_{23} = \pi/4$
and $\delta = -\pi/2$ may naturally hold in the standard
parametrization of $U$ at a given energy scale.
\end{itemize}
It is well known that making use of some proper discrete symmetry groups (such as
$\rm A^{}_4$ and $\rm S^{}_4$) to describe the flavors of charged leptons and
massive neutrinos constitutes the most popular way to achieve $\theta^{}_{23}
= \pi/4$ and $\delta = \pm\pi/2$~\cite{Altarelli:2010gt,Ishimori:2010au,
King:2013eh,Xing:2015fdg,Xing:2019vks,Feruglio:2019ybq,Xing:2022uax}. In
particular, the minimal or residual flavor symmetry of this kind is expected to be
the $\mu$-$\tau$ reflection symmetry in the neutrino sector~\cite{Harrison:2002et}.
A very natural way of obtaining $m^{}_1 = 0$ or $m^{}_3 = 0$ is
to invoke the minimal type-I seesaw model in which two right-handed neutrino
fields are introduced and lepton number violation is required~\cite{Kleppe:1995zz,
Ma:1998zg,Frampton:2002qc}. The so-called Friedberg-Lee
symmetry~\cite{Friedberg:2006it,Lee:2008zzh,Friedberg:2007ba,Friedberg:2007uk,
Friedberg:2009fb,Friedberg:2010zt}, which demands the effective neutrino mass term
to be invariant under the translational transformations
$\nu^{}_\alpha \to \nu^{}_\alpha + \eta^{}_\alpha z^{}_\nu$ with $z^{}_\nu$ being
a constant spinor field and $\eta^{}_\alpha$ denoting a flavor-dependent complex
number (for $\alpha = e, \mu, \tau$), is also an interesting option because it can
not only predict one of the neutrinos to be massless but also help constrain the
flavor texture of massive neutrinos if the values of $\eta^{}_\alpha$ are explicitly
assumed.

So far some attempts have been made to build a simple but viable neutrino mass
model by combining the $\mu$-$\tau$ reflection symmetry with either the minimal
seesaw scenario~\cite{Xing:2006ms,King:2015dvf,King:2016yef,Liu:2017frs,Nath:2018hjx,
Samanta:2017kce,King:2018kka,Nath:2018xih,King:2019tbt,Xing:2020ald,Kang:2021stv,
Zhao:2020bzx,Zhao:2021dwc} or a generalized version of the Friedberg-Lee
symmetry~\cite{Xing:2006xa,Huang:2008ri,Jarlskog:2007qy,He:2009pt,Razzaghi:2012rr,
Zhao:2015bza,Xing:2021zxf}, in order to
simultaneously obtain $m^{}_1 = 0$ (or $m^{}_3 = 0$), $\theta^{}_{23} = \pi/4$ and
$\delta = \pm\pi/2$. Along the latter line of thought, the present paper is
intended to go further and show that such three intriguing predictions can
naturally result from a {\it translational} $\mu$-$\tau$ reflection symmetry
in the neutrino sector; namely, the effective Majorana neutrino mass term keeps
invariant under the following transformations of three left-handed neutrino fields:
\begin{eqnarray}
\nu^{}_{e \rm L} \hspace{-0.2cm} & \to & \hspace{-0.2cm}
(\nu^{}_{e \rm L})^c + U^*_{e i} z^{c}_\nu \; ,
\nonumber \\
\nu^{}_{\mu \rm L} \hspace{-0.2cm} & \to & \hspace{-0.2cm}
(\nu^{}_{\tau \rm L})^c + U^*_{\tau i} z^{c}_\nu \; ,
\nonumber \\
\nu^{}_{\tau \rm L} \hspace{-0.2cm} & \to & \hspace{-0.2cm}
(\nu^{}_{\mu \rm L})^c + U^*_{\mu i} z^{c}_\nu \; ,
\hspace{0.5cm}
\label{eq:2}
%     (2)
\end{eqnarray}
where $U^{}_{\alpha i}$ denote the elements of the $3\times 3$ PMNS flavor mixing
matrix $U$ in its first or third column (i.e., $i = 1$ or $3$), and $z^{c}_\nu$ is
the charge conjugation of a constant spinor field $z^{}_\nu$ which anticommutates
with the neutrino fields~\cite{Xing:2021zxf,Volkov:1973ix,deWit:1975xci}.

As first pointed out in Ref.~\cite{Xing:2021zxf}, a {\it massless} Majorana neutrino
field $\nu^{}_i$ and its charge-conjugated counterpart $\nu^c_i = \nu^{}_i$ satisfy
the same Dirac equation ${\rm i}\gamma^\mu \partial^{}_\mu \nu^{}_i = 0$ which is
invariant under the translational transformation $\nu^{}_i \to \nu^{}_i + z^{}_\nu$
or $\nu^{}_i \to \nu^{c}_i + z^{c}_\nu$ in free space. Such a transformation of
$\nu^{}_i$ is equivalent to
$\nu^{}_{\alpha \rm L} \to \nu^{}_{\alpha \rm L} + U^{}_{\alpha i} z^{}_\nu$ or
$\nu^{}_{\alpha \rm L} \to (\nu^{}_{\alpha \rm L})^c + U^{*}_{\alpha i} z^{c}_\nu$
in the flavor space as a consequence of lepton flavor mixing. This novel observation
clearly explains why the flavor-dependent coefficients $\eta^{}_\alpha$ of
$z^{}_\nu$ in the original Friedberg-Lee transformation or its generalized form
are not really arbitrary but can be uniquely identified as the PMNS matrix elements
$U^{}_{\alpha i}$, and thus it naturally motivates us to conjecture the translational
$\mu$-$\tau$ reflection transformations of $\nu^{}_{\alpha \rm L}$ in
Eq.~(\ref{eq:2}). In this respect our conjecture is certainly new as compared with
those made before (see, e.g., Refs.~\cite{Araki:2009grl,Araki:2009kp,Sinha:2018uop})
and may provide a new angle of view to understand the flavor issues of Majorana
neutrinos. Moreover, we shall extend such a simple working flavor
symmetry to the canonical seesaw framework~\cite{Fritzsch:1975sr,Minkowski:1977sc,
Yanagida:1979as,GellMann:1980vs,Glashow:1979nm,Mohapatra:1979ia}
so as to naturally arrive at a constrained version of the minimal seesaw model.

The remaining parts of this paper are organized as follows. In section 2,
we are going to show that $m^{}_i = 0$ (for $i = 1$ or $3$) will definitely hold if the
effective Majorana neutrino mass term is invariant under the transformations made in
Eq.~(\ref{eq:2}). Section 3 is devoted to extending such a translational
$\mu$-$\tau$ reflection symmetry to the right-handed neutrino sector in the well-known
canonical seesaw mechanism at a superhigh-energy scale, and section 4 is intended
to calculate its soft breaking effects on three neutrino masses, three flavor mixing
angles and two CP-violating phases at the electroweak scale. A brief summary and
some concluding remarks will be made in section 5.

\section{The symmetry and its consequences}

\begin{figure}[t]
\begin{center}
\includegraphics[width=5in]{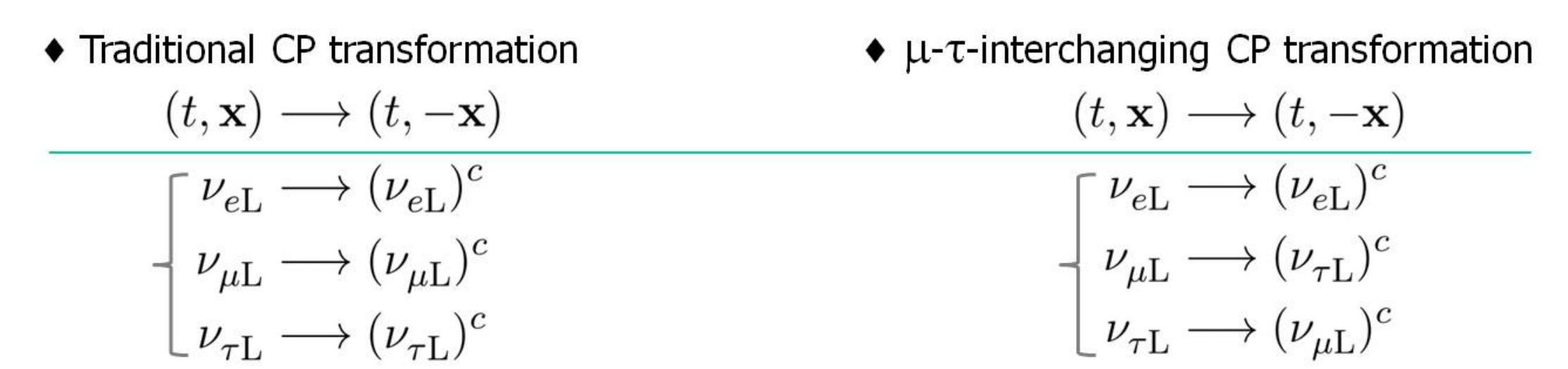}
\vspace{-0.2cm}
\caption{A straightforward comparison between the traditional CP transformation
and the $\mu$-$\tau$-interchanging CP transformation for three left-handed neutrino
fields.}
\label{fig:1}
\end{center}
\end{figure}

Regardless of how the tiny masses of three Majorana neutrinos are generated,
here let us simply focus on their effective mass term ${\cal L}^{}_{\rm mass}$ at
a given energy scale:
\begin{eqnarray}
-{\cal L}^{}_{\rm mass} = \frac{1}{2} \hspace{0.05cm} \overline{\nu^{}_{\rm L}}
\hspace{0.1cm} M^{}_\nu \hspace{0.05cm} (\nu^{}_{\rm L})^c + {\rm h.c.} \; ,
\label{eq:3}
%     (3)
\end{eqnarray}
where $\nu^{}_{\rm L} = (\nu^{}_{e \rm L}, \nu^{}_{\mu \rm L},
\nu^{}_{\tau \rm L})^T$ denotes the column vector of three left-handed
neutrino fields, $(\nu^{}_{\alpha \rm L})^c \equiv {\cal C}
\overline{\nu^{}_{\alpha \rm L}}^T$ stands for the charge conjugate of
$\nu^{}_{\alpha \rm L}$ (for $\alpha = e, \mu, \tau$) with $\cal C$ being the
charge conjugation operator which satisfies ${\cal C}\gamma^{T}_\mu {\cal C}^{-1}
= -\gamma^{}_\mu$, ${\cal C}\gamma^{T}_5 {\cal C}^{-1} = \gamma^{}_5$ and
${\cal C}^{-1} = {\cal C}^\dagger = {\cal C}^{T} = -{\cal C}$, and $M^{}_\nu$
is the $3\times 3$ symmetric Majorana neutrino mass matrix. If the traditional
CP transformations are made for the left-handed neutrino fields, namely
$\nu^{}_{\alpha \rm L} \to (\nu^{}_{\alpha L})^c$ (for $\alpha = e, \mu, \tau$)
as shown in Fig.~\ref{fig:1}, then the invariance of ${\cal L}^{}_{\rm mass}$
requires $M^{}_\nu = M^*_\nu$, implying that this effective Majorana neutrino
mass term must be CP-conserving.

Now let us introduce a real orthogonal $(\mu, \tau)$-associated permutation
matrix, or equivalently one of the six elements of the non-Abelian $\rm S^{}_3$
group, of the form
\begin{eqnarray}
{\cal P} = {\cal P}^T = {\cal P}^\dagger =
\left(\begin{matrix} 1 & 0 & 0 \cr
0 & ~0~ & 1 \cr 0 & 1 & 0 \cr \end{matrix}\right) \; .
\label{eq:4}
%     (4)
\end{eqnarray}
With the help of this notation, we may rewrite the translational $\mu$-$\tau$
reflection transformations proposed in Eq.~(\ref{eq:2}) as follows:
\begin{eqnarray}
\nu^{}_{\rm L} \to {\cal P} \left[(\nu^{}_{\rm L})^c + \xi^*_i z^c_\nu
\right] \; ,
\label{eq:5}
%     (5)
\end{eqnarray}
together with the charge-conjugated transformation
$(\nu^{}_{\rm L})^c \to {\cal P} \left(\nu^{}_{\rm L} + \xi^{}_i z^{}_\nu\right)$,
where $\xi^{}_i$ is a column vector of the PMNS lepton flavor mixing matrix $U$
defined as $\xi^{}_i \equiv (U^{}_{e i} , U^{}_{\mu i} , U^{}_{\tau i})^T$ corresponding
to $\nu^{}_i$ (for $i = 1, 2, 3$). Figure~\ref{fig:1} provides us with a clear
comparison between the traditional CP transformations and the $\mu$-$\tau$-interchanging
CP transformations for three left-handed neutrino fields. Under the latter
transformations the effective Majorana neutrino mass term in Eq.~(\ref{eq:3})
changes to
\begin{eqnarray}
-{\cal L}^{\prime}_{\rm mass} & = & \frac{1}{2} \Big[\overline{\nu^{}_{\rm L}}
\left({\cal P} M^{*}_\nu {\cal P}\right) (\nu^{}_{\rm L})^c +
\overline{\nu^{}_{\rm L}} \left({\cal P} M^{*}_\nu {\cal P}\right) \xi^*_i z^c_\nu
+ \overline{z^{}_\nu} \hspace{0.05cm} \xi^\dagger_i
\left({\cal P} M^{*}_\nu {\cal P}\right)
(\nu^{}_{\rm L})^c
\nonumber \\
&& + \hspace{0.07cm}
\overline{z^{}_\nu} \hspace{0.05cm} \xi^\dagger_i
\left({\cal P} M^{*}_\nu {\cal P}\right) \xi^*_i z^c_\nu\Big] + {\rm h.c.} \; .
\label{eq:6}
%     (6)
\end{eqnarray}
The necessary and sufficient conditions for
${\cal L}^\prime_{\rm mass} = {\cal L}^{}_{\rm mass}$ turn out to be the relation
\begin{eqnarray}
M^{}_\nu = {\cal P} M^{*}_\nu {\cal P} \; ,
\label{eq:7}
%     (7)
\end{eqnarray}
together with three dependent relations
\begin{eqnarray}
M^{}_\nu \xi^*_i = {\bf 0} \; , \quad
\xi^\dagger_i M^{}_\nu = {\bf 0}^T \; , \quad
\xi^\dagger_i M^{}_\nu \xi^*_i = 0 \; ,
\label{eq:8}
%     (8)
\end{eqnarray}
where $\bf 0$ denotes the zero column vector. The condition in Eq.~(\ref{eq:7})
is a natural consequence of the $\mu$-$\tau$ reflection symmetry and provides
a quite strong constraint on the flavor texture of $M^{}_\nu$; and those in
Eq.~(\ref{eq:8}) are apparently required by the translational symmetry of
${\cal L}^{}_{\rm mass}$ and should simply lead us to the result $m^{}_i = 0$.
%%%%%%%%%%%%%%%%%%%%%%%%%%%% Figure 1 %%%%%%%%%%%%%%%%%%%%%%%%%%%%%%%%%%%%%
%%%%%%%%%%%%%%%%%%%%%%%%%%%%%%%%%%%%%%%%%%%%%%%%%%%%%%%%%%%%%%%%%%%%%%%%%%%%

To prove the last point, let us denote the elements of $M^{}_\nu$ as
$\langle m\rangle^{}_{\alpha\beta}$ (for $\alpha, \beta = e, \mu, \tau$)
and express them in terms of the neutrino masses $m^{}_i$ and the PMNS
matrix elements $U^{}_{\alpha i}$ and $U^{}_{\beta i}$ (for $i = 1, 2, 3$)
in the basis where the mass eigenstates of three charged leptons
are identified with their flavor eigenstates. Namely,
$M^{}_\nu = U D^{}_\nu U^T$ with $D^{}_\nu \equiv
{\rm Diag}\left\{m^{}_1, m^{}_2, m^{}_3\right\}$ holds in this flavor basis,
and therefore
\begin{eqnarray}
\left(M^{}_\nu\right)^{}_{\alpha\beta} \equiv
\langle m\rangle^{}_{\alpha\beta} = \sum^3_{i=1}
\left(m^{}_i U^{}_{\alpha i} U^{}_{\beta i} \right) \; .
\label{eq:9}
%     (9)
\end{eqnarray}
With the help of the unitarity of $U$, we combine Eq.~(\ref{eq:8}) with
Eq.~(\ref{eq:9}) and then obtain
%%%%%%%%%%%%%%%%%%%%%%%%%%%%%%%%%%%%%%%%%%%%%%%%%%%%%%%%%%%%%%%%%%%%%%%%%%
\footnote{In Eq.~(\ref{eq:8}) the second condition is actually a
transpose of the first condition, and thus they are equivalent.}
%%%%%%%%%%%%%%%%%%%%%%%%%%%%%%%%%%%%%%%%%%%%%%%%%%%%%%%%%%%%%%%%%%%%%%%%%%
\begin{eqnarray}
&& \sum_\beta \langle m\rangle^{}_{\alpha\beta} U^{*}_{\beta i}
= m^{}_i U^{}_{\alpha i} = 0 \; ,
\nonumber \\
&& \sum_\alpha \sum_\beta U^{*}_{\alpha i} \langle m\rangle^{}_{\alpha\beta}
U^{*}_{\beta i} = m^{}_i = 0 \; .
\hspace{0.5cm}
\label{eq:10}
%     (10)
\end{eqnarray}
So $m^{}_i = 0$ is proved to be another consequence of the translational $\mu$-$\tau$
reflection symmetry of ${\cal L}^{}_{\rm mass}$. Now that $m^{}_2 > m^{}_1$
has been fixed~\cite{Zyla:2020zbs}, we are left with the possibility of
either $m^{}_1 = 0$ (normal ordering) or $m^{}_3 = 0$ (inverted ordering)
for the neutrino mass spectrum. Combining $m^{}_1 = 0$ or $m^{}_3 = 0$ with
the available data on the neutrino mass-squared differences
$\Delta m^2_{ij} \equiv m^2_i - m^2_j$, one may immediately determine the
three neutrino masses. To be more explicit, we have the special normal neutrino
mass hierarchy
\begin{eqnarray}
m^{}_1 \hspace{-0.2cm} & = & \hspace{-0.2cm} 0 \; ,
\nonumber \\
m^{}_2 \hspace{-0.2cm} & = & \hspace{-0.2cm}
\sqrt{\Delta m^2_{21}} = 8.61^{+0.12}_{-0.11} \times 10^{-3}~{\rm eV} \; ,
\nonumber \\
m^{}_3 \hspace{-0.2cm} & = & \hspace{-0.2cm}
\sqrt{\Delta m^2_{31}} = 5.01^{+0.03}_{-0.03} \times 10^{-2}~{\rm eV} \; ,
\hspace{0.5cm}
\label{eq:11}
%     (11)
\end{eqnarray}
where $\Delta m^2_{21} = 7.42^{+0.21}_{-0.20} \times 10^{-5}~{\rm eV}^2$ and
$\Delta m^2_{31} = 2.510^{+0.027}_{-0.027} \times 10^{-3}~{\rm eV}^2$ as the
best-fit values plus the $\pm 1\sigma$
ranges~\cite{Capozzi:2021fjo,Gonzalez-Garcia:2021dve} have been input;
or the special inverted neutrino mass hierarchy
\begin{eqnarray}
m^{}_1 \hspace{-0.2cm} & = & \hspace{-0.2cm}
\sqrt{\left|\Delta m^2_{31}\right|} = 4.92^{+0.03}_{_-0.03}
\times 10^{-2}~{\rm eV} \; ,
\nonumber \\
m^{}_2 \hspace{-0.2cm} & = & \hspace{-0.2cm}
\sqrt{\left|\Delta m^2_{32}\right|} = 4.99^{+0.03}_{-0.03}
\times 10^{-2}~{\rm eV} \; , \hspace{0.5cm}
\nonumber \\
m^{}_3 \hspace{-0.2cm} & = & \hspace{-0.2cm} 0 \; ,
\label{eq:12}
%     (12)
\end{eqnarray}
where $\Delta m^2_{21} = 7.42^{+0.21}_{-0.20} \times 10^{-5}~{\rm eV}^2$ and
$\Delta m^2_{32} = -2.490^{+0.026}_{-0.028} \times
10^{-3}~{\rm eV}^2$~\cite{Capozzi:2021fjo,Gonzalez-Garcia:2021dve}
have been used for a numerical illustration.
The next-generation neutrino oscillation experiments are going to determine
which neutrino mass ordering is really true, and the next-generation precision
measurements of the cosmic microwave background anisotropies and large scale
structures may hopefully help to probe the absolute neutrino mass scale and
examine whether $m^{}_1$ or $m^{}_3$ is essentially vanishing or vanishingly
small~\cite{Xing:2019vks}.

No matter whether $m^{}_1 = 0$ or $m^{}_3 = 0$ holds, we are left with only
two nontrivial CP-violating phases $\delta$ and $\sigma$ in the PMNS matrix
$U$~\cite{Xing:2020ald}:
\begin{eqnarray}
U = P^{}_l \left(\begin{matrix}
c^{}_{12} c^{}_{13} & s^{}_{12} c^{}_{13} & s^{}_{13} e^{-{\rm i} \delta} \cr
-s^{}_{12} c^{}_{23} - c^{}_{12} s^{}_{13} s^{}_{23}  e^{{\rm i} \delta}
& c^{}_{12} c^{}_{23} - s^{}_{12} s^{}_{13} s^{}_{23} e^{{\rm i} \delta}
& c^{}_{13} s^{}_{23} \cr
s^{}_{12} s^{}_{23} - c^{}_{12} s^{}_{13} c^{}_{23}  e^{{\rm i} \delta}
& ~ -c^{}_{12} s^{}_{23} - s^{}_{12} s^{}_{13} c^{}_{23} e^{{\rm i} \delta} ~
& c^{}_{13} c^{}_{23} \cr\end{matrix}\right) P^{}_\nu \; ,
\label{eq:13}
%     (13)
\end{eqnarray}
where $P^{}_l = {\rm Diag}\left\{e^{{\rm i}\phi^{}_e}, e^{{\rm i}\phi^{}_\mu},
e^{{\rm i}\phi^{}_\tau}\right\}$ has no physical effects, but
$P^{}_\nu = {\rm Diag}\left\{1, e^{{\rm i}\sigma}, 1\right\}$ with $\sigma$
being the Majorana CP phase is sensitive to lepton number violation.
Substituting Eq.~(\ref{eq:13}) into Eq.~(\ref{eq:9}) and then taking account
of
\begin{eqnarray}
\langle m\rangle^{}_{ee} = \langle m\rangle^{*}_{ee} \; , \quad
\langle m\rangle^{}_{\mu\tau} = \langle m\rangle^{*}_{\mu\tau} \; , \quad
\langle m\rangle^{}_{e\tau} = \langle m\rangle^{*}_{e\mu} \; , \quad
\langle m\rangle^{}_{\tau\tau} = \langle m\rangle^{*}_{\mu\mu} \;
\label{eq:14}
%     (14)
\end{eqnarray}
derived from Eq.~(\ref{eq:7}), we simply arrive at
\begin{eqnarray}
\theta^{}_{23} = \frac{\pi}{4} \; , \quad
\delta = \pm\frac{\pi}{2} \; , \quad
\sigma = 0 ~~{\rm or}~~ \frac{\pi}{2} \;
\label{eq:15}
%     (15)
\end{eqnarray}
as a natural consequence of the $\mu$-$\tau$ reflection symmetry no matter
whether $m^{}_1 = 0$ or $m^{}_3 = 0$ holds, together with the conditions
$\phi^{}_e = 0$ or $\pm\pi$ and $\phi^{}_\mu + \phi^{}_\tau = 0$ or $\pi$.
The preliminary T2K measurement favors the possibility of
$\delta = -\pi/2$~\cite{T2K:2019bcf}. At present the measurement of a
neutrinoless double-beta ($0\nu 2\beta$) decay mode, whose half-life is sensitive
to the magnitude of $\langle m\rangle^{}_{ee}$, is the only experimentally feasible
way to probe the Majorana nature of massive neutrinos and constrain the
corresponding phase parameters. In the above translational $\mu$-$\tau$
reflection symmetry limit, we have
\begin{eqnarray}
\left|\langle m\rangle^{}_{ee}\right|^{}_{m^{}_1 = 0}
\hspace{-0.2cm} & = & \hspace{-0.2cm}
m^{}_2 s^2_{12} c^2_{13} \mp m^{}_3 s^2_{13}
\simeq \left\{\begin{array}{l}
1.45 \times 10^{-3} ~{\rm eV} \quad\quad (\sigma = 0) \; , \\
\vspace{-0.6cm} \\
3.67 \times 10^{-3} ~{\rm eV} \quad\quad (\sigma = \pi/2) \; ; \end{array} \right.
\nonumber \\
\left|\langle m\rangle^{}_{ee}\right|^{}_{m^{}_3 = 0}
\hspace{-0.2cm} & = & \hspace{-0.2cm}
c^2_{13} \left(m^{}_1 c^2_{12} \pm m^{}_2 s^2_{12} \right)
\simeq \left\{\begin{array}{l}
4.83 \times 10^{-2} ~{\rm eV} \quad\quad (\sigma = 0) \; , \\
\vspace{-0.6cm} \\
1.91 \times 10^{-2} ~{\rm eV} \quad\quad (\sigma = \pi/2) \; , \end{array} \right.
\hspace{0.5cm}
\label{eq:16}
%     (16)
\end{eqnarray}
where the best-fit values $s^{2}_{12} = 0.304$ and $s^2_{13} = 0.022$
~\cite{Capozzi:2021fjo,Gonzalez-Garcia:2021dve} together with the numerical
results of $m^{}_i$ (for $i = 1, 2, 3$) obtained in Eqs.~(\ref{eq:11}) and
(\ref{eq:12}) have been used as the typical inputs. Eq.~(\ref{eq:16}) may give
one a ball-park but promising feeling that the translational $\mu$-$\tau$
reflection symmetry is so powerful that it can really lead us to a definite
prediction for the effective electron-neutrino mass of the $0\nu 2\beta$ decays.
In this special case an experimental measurement of
$\left|\langle m\rangle^{}_{ee}\right|$ will allow us not only to determine
the neutrino mass ordering but also to pin down the Majorana CP-violating phase
$\sigma$.

As a by-product, the other three independent matrix elements of $M^{}_\nu$
in the $\mu$-$\tau$ reflection symmetry limit can similarly be determined.
We have the results as follows:
\begin{eqnarray}
\left|\langle m\rangle^{}_{\mu\mu}\right|^{}_{m^{}_1 = 0}
\hspace{-0.2cm} & = & \hspace{-0.2cm}
\frac{1}{2} \sqrt{\displaystyle
\left[ m^{}_2 \left( c^2_{12} - s^2_{12} s^2_{13}\right)
\pm m^{}_3 c^2_{13}\right]^2 + 4 m^2_2 c^2_{12} s^2_{12} s^2_{13}}
\nonumber \\
\hspace{-0.2cm} & \simeq & \hspace{-0.2cm}
\left\{\begin{array}{l}
2.75 \times 10^{-2} ~{\rm eV} \quad\quad (\sigma = 0) \; , \\
\vspace{-0.6cm} \\
2.15 \times 10^{-2} ~{\rm eV} \quad\quad (\sigma = \pi/2) \; ; \end{array} \right.
\nonumber \\
\left|\langle m\rangle^{}_{\mu\mu}\right|^{}_{m^{}_3 = 0}
\hspace{-0.2cm} & = & \hspace{-0.2cm}
\frac{1}{2} \sqrt{\displaystyle \left[ m^{}_1 \left( s^2_{12} -
c^2_{12} s^2_{13} \right) \pm m^{}_2 \left( c^2_{12} - s^2_{12} s^2_{13}
\right)\right]^2 + 4 \left(m^{}_1 \mp m^{}_2\right)^2 c^2_{12} s^2_{12}
s^2_{13}} \hspace{0.5cm}
\nonumber \\
\hspace{-0.2cm} & \simeq & \hspace{-0.2cm}
\left\{\begin{array}{l}
2.43 \times 10^{-2} ~{\rm eV} \quad\quad (\sigma = 0) \; , \\
\vspace{-0.6cm} \\
1.22 \times 10^{-2} ~{\rm eV} \quad\quad (\sigma = \pi/2) \; ; \end{array} \right.
\label{eq:17}
%     (17)
\end{eqnarray}
and
\begin{eqnarray}
\left|\langle m\rangle^{}_{e\mu}\right|^{}_{m^{}_1 = 0}
\hspace{-0.2cm} & = & \hspace{-0.2cm}
\frac{c^{}_{13}}{\sqrt 2} \sqrt{\displaystyle
m^{2}_2 c^2_{12} s^2_{12} + \left(m^{}_2 s^2_{12} \pm m^{}_3 \right)^2 s^2_{13}}
\nonumber \\
\hspace{-0.2cm} & \simeq & \hspace{-0.2cm}
\left\{\begin{array}{l}
6.13 \times 10^{-3} ~{\rm eV} \quad\quad (\sigma = 0) \; , \\
\vspace{-0.6cm} \\
5.65 \times 10^{-3} ~{\rm eV} \quad\quad (\sigma = \pi/2) \; ; \end{array} \right.
\nonumber \\
\left|\langle m\rangle^{}_{e\mu}\right|^{}_{m^{}_3 = 0}
\hspace{-0.2cm} & = & \hspace{-0.2cm}
\frac{c^{}_{13}}{\sqrt 2} \sqrt{\displaystyle
\left( m^{}_1 \mp m^{}_2 \right)^2 c^2_{12} s^2_{12} +
\left( m^{}_1 c^2_{12} \pm m^{}_2 s^2_{12}\right)^2 s^2_{13}} \hspace{0.5cm}
\nonumber \\
\hspace{-0.2cm} & \simeq & \hspace{-0.2cm}
\left\{\begin{array}{l}
5.13 \times 10^{-3} ~{\rm eV} \quad\quad (\sigma = 0) \; , \\
\vspace{-0.6cm} \\
3.19 \times 10^{-2} ~{\rm eV} \quad\quad (\sigma = \pi/2) \; ; \end{array} \right.
\label{eq:18}
%     (18)
\end{eqnarray}
as well as
\begin{eqnarray}
\left|\langle m\rangle^{}_{\mu\tau}\right|^{}_{m^{}_1 = 0}
\hspace{-0.2cm} & = & \hspace{-0.2cm}
\frac{1}{2} \left[ m^{}_3 c^2_{13} \mp m^{}_2 \left( c^2_{12} +
s^{2}_{12} s^2_{13}\right)\right]
\nonumber \\
\hspace{-0.2cm} & \simeq & \hspace{-0.2cm}
\left\{\begin{array}{l}
2.15 \times 10^{-2} ~{\rm eV} \quad\quad (\sigma = 0) \; , \\
\vspace{-0.6cm} \\
2.75 \times 10^{-2} ~{\rm eV} \quad\quad (\sigma = \pi/2) \; ; \end{array} \right.
\nonumber \\
\left|\langle m\rangle^{}_{\mu\tau}\right|^{}_{m^{}_3 = 0}
\hspace{-0.2cm} & = & \hspace{-0.2cm}
\frac{1}{2} \left[ m^{}_2 \left( c^2_{12} + s^2_{12} s^2_{13} \right)
\pm m^{}_1 \left( s^2_{12} + c^2_{12} s^2_{13}\right)\right] \hspace{0.5cm}
\nonumber \\
\hspace{-0.2cm} & \simeq & \hspace{-0.2cm}
\left\{\begin{array}{l}
2.54 \times 10^{-2} ~{\rm eV} \quad\quad (\sigma = 0) \; , \\
\vspace{-0.6cm} \\
9.68 \times 10^{-3} ~{\rm eV} \quad\quad (\sigma = \pi/2) \; . \end{array} \right.
\label{eq:19}
%     (19)
\end{eqnarray}
Note that all the results in Eqs.~(\ref{eq:16})---(\ref{eq:19}) are insensitive
to the two-fold uncertainties of $\delta$ (i.e., $\delta = \pm \pi/2$) in the
translational $\mu$-$\tau$ reflection symmetry limit. The smallness of these matrix
elements makes it extremely difficult, if not impossible, to probe the relevant
lepton-number-violating processes mediated by $\nu^{}_i$ (for $i = 1, 2, 3$),
as such reactions are directly associated with the magnitudes of
$\langle m\rangle^{}_{\alpha\beta}$~\cite{Rodejohann:2011mu,Bilenky:2014uka,
Dolinski:2019nrj}.

\begin{figure}[t]
\begin{center}
\includegraphics[width=4.4in]{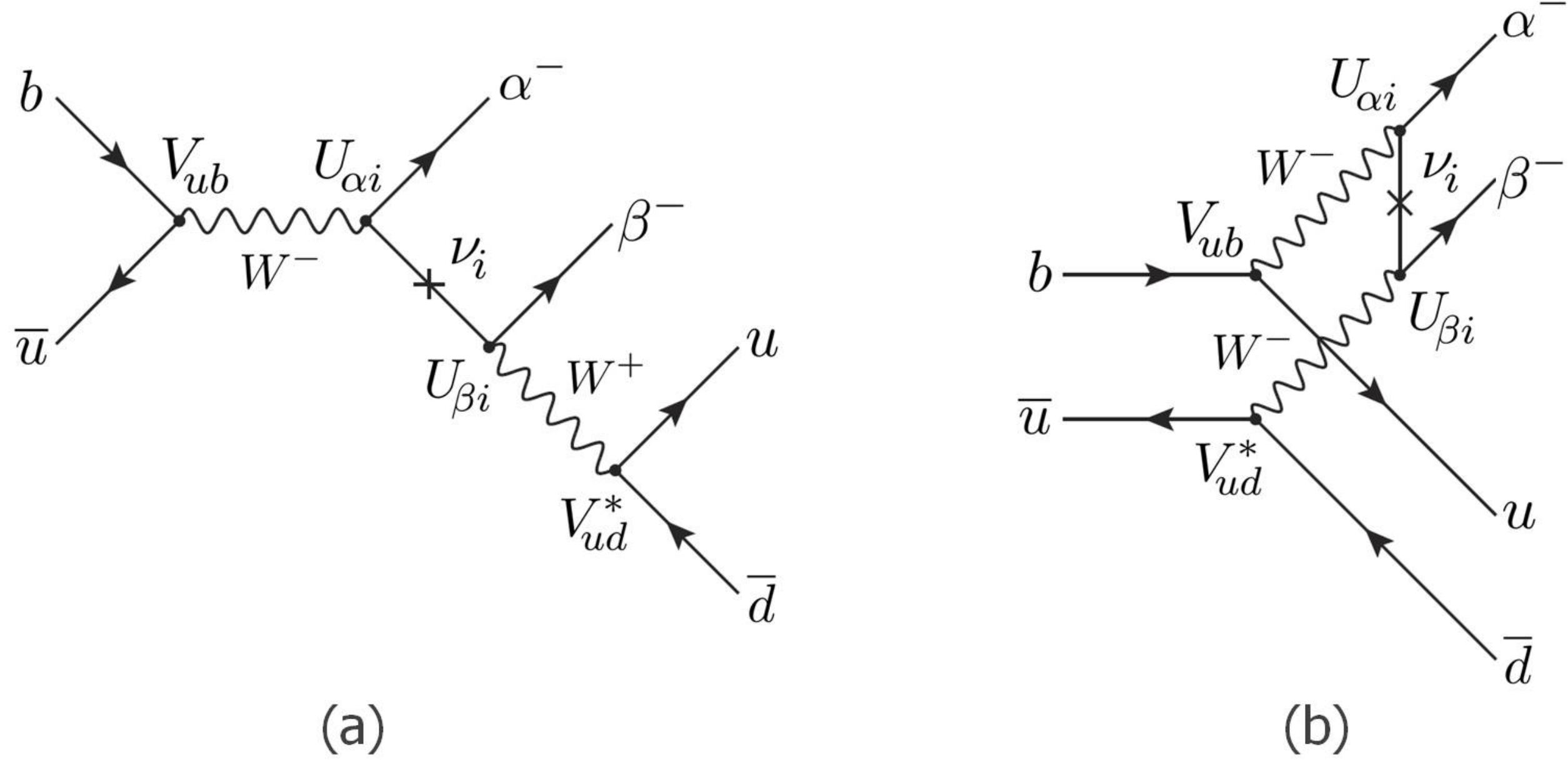}
\vspace{-0.25cm}
\caption{The Feynman diagrams for a lepton-number-violating decay mode
$B^-_u \to \pi^+ \alpha^- \beta^-$ (for $\alpha, \beta = e, \mu, \tau$),
where $\nu^{}_i$ (for $i = 1, 2, 3$) denote the neutrino mass eigenstates,
$U^{}_{\alpha i}$ and $U^{}_{\beta i}$ are the PMNS matrix elements,
$V^{}_{ud}$ and $V^{}_{ub}$ are the CKM matrix elements.}
\label{fig:2}
\end{center}
\end{figure}

For the sake of illustration,
here let us briefly comment on a typical example of this kind:
the lepton-number-violating decay modes $B^-_u \to \pi^+ \alpha^- \beta^-$
which involve all the six effective Majorana neutrino masses
$\langle m\rangle^{}_{\alpha\beta}$ (for $\alpha, \beta = e, \mu, \tau$).
The two tree-level Feynman diagrams
responsible for $B^-_u \to \pi^+ \alpha^- \beta^-$ are shown in Fig.~\ref{fig:2}.
Quite similar to the more familiar case of those $0\nu 2\beta$
processes which are directly associated with $|\langle m\rangle^{}_{ee}|^2$,
the decay rates $\Gamma (B^-_u \to \pi^+ \alpha^- \beta^-)$ are
simply proportional to $|\langle m\rangle^{}_{\alpha\beta}|^2$. So the
$\mu$-$\tau$ reflection symmetry under discussion allows us to obtain
$\Gamma (B^-_u \to \pi^+ e^- \tau^-) = \Gamma (B^-_u \to \pi^+ e^- \mu^-)$ and
$\Gamma (B^-_u \to \pi^+ \tau^- \tau^-) = \Gamma (B^-_u \to \pi^+ \mu^- \mu^-)$.
But there are only very preliminary upper bounds on the branching ratios
of $B^-_u \to \pi^+ e^- e^-$, $B^-_u \to \pi^+ e^- \mu^-$ and
$B^-_u \to \pi^+ \mu^- \mu^-$~\cite{Zyla:2020zbs} which cannot provide
any meaningful constraints on $|\langle m\rangle^{}_{ee}|$,
$|\langle m\rangle^{}_{e\mu}|$ and $|\langle m\rangle^{}_{\mu\mu}|$ as
compared with the more realistic and feasible $0\nu 2\beta$-decay experiments.
This situation is expected to be improved to some extent in the upcoming precision
measurement era characterized by the LHCb experiment at the high-luminosity
LHC~\cite{LHCb:2018roe} and the Belle-II experiments at the KEK super-$B$
factory~\cite{Belle-II:2018jsg}.
%%%%%%%%%%%%%%%%%%%%%%%%%%%% Figure 2 %%%%%%%%%%%%%%%%%%%%%%%%%%%%%%%%%%%%%%%%%%%%
%%%%%%%%%%%%%%%%%%%%%%%%%%%%%%%%%%%%%%%%%%%%%%%%%%%%%%%%%%%%%%%%%%%%%%%%%%%%%%%%%

So far we have simply assumed the translational $\mu$-$\tau$ reflection symmetry
of three active Majorana neutrinos at low energies. It is certainly
appropriate to conjecture such a working flavor symmetry at a superhigh-energy
scale where there may exist a well-defined mechanism responsible for the origin
of tiny neutrino masses and a suitable flavor symmetry to constrain the
neutrino mass texture. In this regard the popular canonical seesaw
mechanism~\cite{Fritzsch:1975sr,Minkowski:1977sc,Yanagida:1979as,GellMann:1980vs,
Glashow:1979nm,Mohapatra:1979ia}, which attributes the smallness of three active
neutrino masses to the existence of three heavy Majorana neutrinos and the
violation of lepton number conservation, is no doubt the most
natural theoretical framework at hand.

\section{An extension in the seesaw framework}

The canonical seesaw mechanism is a straightforward and natural extension of the
SM with three right-handed neutrino fields $N^{}_{\alpha \rm R}$ (for
$\alpha = e, \mu, \tau$) and lepton number violation~\cite{Fritzsch:1975sr,
Minkowski:1977sc,Yanagida:1979as,GellMann:1980vs,Glashow:1979nm,Mohapatra:1979ia}.
In this theoretical framework the gauge- and Lorentz-invariant neutrino mass terms
can be written as
\begin{eqnarray}
-{\cal L}^{}_\nu = \overline{\ell^{}_{\rm L}} \hspace{0.05cm} Y^{}_\nu
\widetilde{H} N^{}_{\rm R} + \frac{1}{2} \hspace{0.05cm} \overline{(N^{}_{\rm R})^c}
\hspace{0.05cm} M^{}_{\rm R} N^{}_{\rm R} + {\rm h.c.} \; ,
\label{20}
%     (20)
\end{eqnarray}
where $\ell^{}_{\rm L}$ denotes the $\rm SU(2)^{}_{\rm L}$ doublet of the
left-handed lepton fields, $\widetilde{H} \equiv {\rm i} \sigma^{}_2 H^*$ with
$H$ being the Higgs doublet and $\sigma^{}_2$ being the second Pauli matrix,
$N^{}_{\rm R} = (N^{}_{e \rm R} , N^{}_{\mu \rm R} , N^{}_{\tau \rm R})^T$
is the column vector of three right-handed neutrino fields which are the
$\rm SU(2)^{}_{\rm L}$ singlets,
$(N^{}_{\rm R})^c \equiv {\cal C} \overline{N^{}_{\rm R}}^T$ is the charge
conjugate of $N^{}_{\alpha \rm R}$, $Y^{}_\nu$ represents an arbitrary $3\times 3$
Yukawa coupling matrix, and $M^{}_{\rm R}$ stands for a symmetric $3\times 3$
Majorana mass matrix. After spontaneous electroweak gauge symmetry breaking,
Eq.~(\ref{20}) becomes
\begin{eqnarray}
-\widetilde{\cal L}^{}_{\rm mass} = \overline{\nu^{}_{\rm L}} \hspace{0.05cm}
M^{}_{\rm D} N^{}_{\rm R} + \frac{1}{2} \hspace{0.05cm} \overline{(N^{}_{\rm R})^c}
\hspace{0.05cm} M^{}_{\rm R} N^{}_{\rm R} + {\rm h.c.} \; ,
\label{eq:21}
%     (21)
\end{eqnarray}
where $\nu^{}_{\rm L}$ is the column vector of $\nu^{}_{e \rm L}$,
$\nu^{}_{\mu \rm L}$ and $\nu^{}_{\tau \rm L}$ as already given in Eq.~(\ref{eq:3}),
and $M^{}_{\rm D} \equiv Y^{}_\nu \langle H\rangle$ with
$\langle H\rangle \simeq 174~{\rm GeV}$ being
the vacuum expectation value of the Higgs field. Note that $M^{}_{\rm D}$ is
in general neither Hermitian nor symmetric. In a way similar to Eq.~(\ref{eq:5}),
we may make the following transformations for the left- and right-handed neutrino
fields
%%%%%%%%%%%%%%%%%%%%%%%%%%%%%%%%%%%%%%%%%%%%%%%%%%%%%%%%%%%%%%%%%%%%%%%%%%%%%%%%
\footnote{Here we only make the $\mu$-$\tau$ reflection transformation instead
of the translational $\mu$-$\tau$ reflection transformation for the left-handed
neutrino fields. Otherwise, the constraint on $M^{}_{\rm D}$ would be too strong
to result in a viable texture for the effective Majorana neutrino mass matrix
$M^{}_\nu$. If the $\mu$-$\tau$ reflection transformation $\nu^{}_{\rm L}
\to {\cal P} (\nu^{}_{\rm L})^c$ were not made, on the other hand, the constraint
on $M^{}_\nu$ would be too weak to be desirable for our purpose of enhancing
the predictability of this model.}:
%%%%%%%%%%%%%%%%%%%%%%%%%%%%%%%%%%%%%%%%%%%%%%%%%%%%%%%%%%%%%%%%%%%%%%%%%%%%%%%%
\begin{eqnarray}
\nu^{}_{\rm L} \to {\cal P} (\nu^{}_{\rm L})^c \; , \quad
N^{}_{\rm R} \to {\cal P} \left[(N^{}_{\rm R})^c + \zeta^*_i z^c_N
\right] \; ,
\label{eq:22}
%     (22)
\end{eqnarray}
together with $(\nu^{}_{\rm L})^c \to {\cal P} \nu^{}_{\rm L}$
and $(N^{}_{\rm R})^c \to {\cal P} \left(N^{}_{\rm R} + \zeta^{}_i z^{}_N\right)$,
where the $(\mu, \tau)$-associated permutation matrix $\cal P$ has been given in
Eq.~(\ref{eq:4}), $\zeta^{}_i$ is a column vector of the $3\times 3$ unitary flavor
mixing matrix $U^\prime$ used to diagonalize $M^{}_{\rm R}$ and defined as
$\zeta^{}_i \equiv (U^{\prime}_{e i} , U^{\prime}_{\mu i} , U^{\prime}_{\tau i})^T$
corresponding to the heavy neutrino mass eigenstate $N^{}_i$ with the mass $M^{}_i$
(for $i = 1, 2, 3$), and $z^{}_N$ is another constant spinor field like $z^{}_\nu$
in Eq.~(\ref{eq:2}). Then Eq.~(\ref{eq:21}) becomes
\begin{eqnarray}
-\widetilde{\cal L}^{\prime}_{\rm mass} & = & \overline{\nu^{}_{\rm L}} \hspace{0.05cm}
\left({\cal P} M^{*}_{\rm D} {\cal P}\right) N^{}_{\rm R} + \frac{1}{2} \Big[
\overline{(N^{}_{\rm R})^c} \left({\cal P} M^{*}_{\rm R} {\cal P}\right)
N^{}_{\rm R} + \overline{(N^{}_{\rm R})^c} \left({\cal P} M^{*}_{\rm R} {\cal P}\right)
\zeta^{}_i z^{}_N
\nonumber \\
&& + \hspace{0.1cm} \overline{z^c_N} \hspace{0.1cm} \zeta^T_i
\left({\cal P} M^{*}_{\rm R} {\cal P}\right) N^{}_{\rm R} +
\overline{z^c_N} \hspace{0.1cm} \zeta^T_i \left({\cal P} M^{*}_{\rm R} {\cal P}\right)
\zeta^{}_i z^{}_N \Big] + {\rm h.c.} \; .
\label{eq:23}
%     (23)
\end{eqnarray}
In this case the necessary and sufficient conditions for
$\widetilde{\cal L}^{\prime}_{\rm mass} = \widetilde{\cal L}^{}_{\rm mass}$
turn out to be
\begin{eqnarray}
M^{}_{\rm D} = {\cal P} M^{*}_{\rm D} {\cal P} \; , \quad
M^{}_{\rm R} = {\cal P} M^{*}_{\rm R} {\cal P} \; ,
\label{eq:24}
%     (24)
\end{eqnarray}
together with the dependent relations
\begin{eqnarray}
M^{}_{\rm D} \hspace{0.05cm} \zeta^{}_i = {\bf 0} \; , \quad
M^{}_{\rm R} \hspace{0.05cm} \zeta^{}_i = {\bf 0} \; , \quad
\zeta^T_i M^{}_{\rm R} = {\bf 0}^T \; , \quad
\zeta^T_i M^{}_{\rm R} \hspace{0.05cm} \zeta^{}_i = 0 \; ,
\label{eq:25}
%     (25)
\end{eqnarray}
where $\bf 0$ stands for the zero column vector. The flavor textures of
$M^{}_{\rm D}$ and $M^{}_{\rm R}$ constrained by Eq.~(\ref{eq:24}) can
therefore be expressed as follows~\cite{Xing:2017mkx,Xing:2019edp}:
\begin{eqnarray}
M^{}_{\rm D} \hspace{-0.2cm} & = & \hspace{-0.2cm} \left(\begin{matrix}
a & b & b^* \cr
e & c & d \cr
e^* & ~~d^*~~ & c^* \end{matrix}\right) \; ,
\nonumber \\
M^{}_{\rm R} \hspace{-0.2cm} & = & \hspace{-0.2cm} \left(\begin{matrix}
A & B & B^* \cr
B & C & D \cr
B^* & D & C^* \end{matrix}\right) \; ,
\hspace{0.5cm}
\label{eq:26}
%     (26)
\end{eqnarray}
where the matrix elements $a$, $A$ and $D$ are real. On the other hand,
Eq.~(\ref{eq:25}) implies that $M^{}_i = 0$ must hold. To assure the seesaw
mechanism to work, however, only one of the three mass eigenstates $N^{}_i$
(for $i = 1, 2, 3$) is allowed to vanish.

Without loss of any generality, let us take $M^{}_1 = 0$. Now that there is
no way to measure $U^\prime$ in any realistic experiments, we simply choose
the flavor basis defined by $\zeta^{}_1 \equiv (1, 0, 0)^T$. Then
Eq.~(\ref{eq:25}) immediately leads us to
\begin{eqnarray}
M^{}_{\rm D} \hspace{-0.2cm} & = & \hspace{-0.2cm} \left(\begin{matrix}
0 & b & b^* \cr
0 & c & d \cr
0 & ~~d^*~~ & c^* \end{matrix}\right) \; ,
\nonumber \\
M^{}_{\rm R} \hspace{-0.2cm} & = & \hspace{-0.2cm} \left(\begin{matrix}
0 & 0 & 0 \cr
0 & C & D \cr
0 & \hspace{0.1cm} D \hspace{0.1cm} & C^* \end{matrix}\right) \; .
\hspace{0.8cm}
\label{eq:27}
%     (27)
\end{eqnarray}
As argued in Ref.~\cite{Jarlskog:2007qy}, the flavor eigenstate of the massless
neutrino $N^{}_1$ simply decouples and has only gravitational interactions due
to its kinetic energy term. We actually have no idea about whether such a massless
sterile neutrino state could really exist in nature. Here our focus is on the
point that the translational $\mu$-$\tau$ reflection symmetry of
$\widetilde{\cal L}^{}_{\rm mass}$ can be regarded as a reasonable
way to reduce the degrees of freedom associated with the canonical seesaw
mechanism and constrain its flavor textures. Through this purely phenomenological
operation, we only need to consider the remaining two right-handed neutrino
fields and may simply rewrite $\widetilde{\cal L}^{}_{\rm mass}$ in
Eq.~(\ref{eq:21}) as
\begin{eqnarray}
-\widehat{\cal L}^{}_{\rm mass} = \overline{\nu^{}_{\rm L}} \hspace{0.05cm}
\widehat{M}^{}_{\rm D} N^{\prime}_{\rm R} + \frac{1}{2} \hspace{0.05cm}
\overline{(N^{\prime}_{\rm R})^c}
\hspace{0.05cm} \widehat{M}^{}_{\rm R} N^{\prime}_{\rm R} + {\rm h.c.} \; ,
\label{eq:28}
%     (28)
\end{eqnarray}
where $N^{\prime}_{\rm R} \equiv (N^{}_{\mu \rm R} , N^{}_{\tau \rm R})^T$ is
defined, and the effective neutrino mass matrices $\widehat{M}^{}_{\rm D}$ and
$\widehat{M}^{}_{\rm R}$ are of the following textures:
\begin{eqnarray}
\widehat{M}^{}_{\rm D} = \left(\begin{matrix}
b & ~b^* \cr
c & ~d \cr
d^* & ~c^* \end{matrix}\right) \; , \quad
\widehat{M}^{}_{\rm R} = \left(\begin{matrix}
C & ~D \cr
D & ~C^* \end{matrix}\right) \; ,
\label{eq:29}
%     (29)
\end{eqnarray}
which assure $\widehat{\cal L}^{}_{\rm mass}$ itself to satisfy the $\mu$-$\tau$
reflection symmetry. Such a scenario was first proposed in the minimal seesaw
framework~\cite{Nath:2018hjx}, and here it is reproduced from the canonical seesaw
mechanism with the help of the translational $\mu$-$\tau$ reflection symmetry.

Given Eqs.~(\ref{eq:28}) and (\ref{eq:29}), in which the heavy degrees of freedom
can be integrated out to obtain the unique dimension-five Weinberg operator for
three active Majorana neutrino fields~\cite{Weinberg:1979sa}, the resulting
effective Majorana neutrino mass matrix
\begin{eqnarray}
\widehat{M}^{}_\nu \simeq -\widehat{M}^{}_{\rm D} \widehat{M}^{-1}_{\rm R}
\widehat{M}^{T}_{\rm D} \;
\label{eq:30}
%     (30)
\end{eqnarray}
must be of rank two and thus have a zero eigenvalue (i.e., either $m^{}_1 = 0$
or $m^{}_3 = 0$)
%%%%%%%%%%%%%%%%%%%%%%%%%%%%%%%%%%%%%%%%%%%%%%%%%%%%%%%%%%%%%%%%%%%%%%%%%%%%%%%%%%%
\footnote{Note that this observation is a natural consequence of the ``seesaw fair
play rule"~\cite{Xing:2007uq} which is independent of the approximation made
in obtaining the seesaw formula in Eq.~(\ref{eq:30}).}.
%%%%%%%%%%%%%%%%%%%%%%%%%%%%%%%%%%%%%%%%%%%%%%%%%%%%%%%%%%%%%%%%%%%%%%%%%%%%%%%%%%%
Substituting Eq.~(\ref{eq:29}) into Eq.~(\ref{eq:30}), one may easily
show that $\widehat{M}^{}_\nu$ satisfies
\begin{eqnarray}
\widehat{M}^{}_\nu = {\cal P} \widehat{M}^{*}_\nu {\cal P} \; ,
\label{eq:31}
%     (31)
\end{eqnarray}
just as $M^{}_\nu$ in Eq.~(\ref{eq:7}). So the relevant matrix elements
and flavor mixing parameters of $\widehat{M}^{}_\nu$ are constrained as in
Eqs.~(\ref{eq:14}) and (\ref{eq:15}).

Now that $\widehat{M}^{}_\nu$ is obtained at a superhigh-energy scale
$\Lambda^{}_{\mu\tau}$ where both the minimal seesaw mechanism and the translational
$\mu$-$\tau$ reflection symmetry take effect, it will unavoidably receive some
quantum corrections at the electroweak scale $\Lambda^{}_{\rm EW} \sim 10^2~{\rm GeV}$
through the renormalization-group-equation (RGE) running effects. At the one-loop level
such RGE-induced quantum corrections provide a very simple and natural way of softly
breaking the exact $\mu$-$\tau$ reflection symmetry but keeping one of the three active
neutrinos massless. Let us proceed to elaborate on this point.

\section{RGE-induced soft symmetry breaking}

The RGE approach, which was originally invented and developed in the
1950s~\cite{StueckelbergdeBreidenbach:1952pwl,Gell-Mann:1954yli,Bogolyubov:1956gh},
is a very powerful theoretical tool. It has proved to be staggeringly successful in the
study of critical phenomena in condensed matter physics~\cite{Wilson:1971bg,Wilson:1971dh}
and in the discovery of asymptotic freedom of strong interactions in quantum field theories
and particle physics~\cite{Gross:1973id,Politzer:1973fx}. The key point of the RGE tool is
that the physical parameters at one renormalization point or energy scale can be
related to their counterparts at another renormalization point or energy scale with
the help of the relevant differential equations, such that the theory keeps its form
invariance or self similarity between the two points or scales. Regarding the flavor
issues of charged fermions and massive neutrinos, one is especially interested
in the RGE-induced connection between a superhigh energy scale characterized by a
fundamental mechanism of neutrino mass generation and the electroweak scale where the
physical flavor parameters can be experimentally measured~\cite{Xing:2019vks,Ohlsson:2013xva}.

The one-loop differential RGE for the effective Majorana neutrino mass matrix
is well known to us~\cite{Chankowski:1993tx,Babu:1993qv,Casas:1999tg,Antusch:2001ck},
and its integral form can also be found in the literature
(see, e.g., Refs.~\cite{Ellis:1999my,Fritzsch:1999ee,Mei:2003gn,Zhang:2020lsd}).
Given $M^{}_\nu (\Lambda^{}_{\mu\tau})$ at the $\mu$-$\tau$ reflection symmetry
scale and in the basis where the mass and flavor eigenstates of three charged
leptons are chosen to be identical, its counterpart $M^{}_\nu (\Lambda^{}_{\rm EW})$
at the electroweak scale is explicitly expressed as
\begin{eqnarray}
M^{}_\nu (\Lambda^{}_{\rm EW}) =
I^{}_\nu (\Lambda^{}_{\rm EW}) \left[T^{}_l (\Lambda^{}_{\rm EW})
\cdot M^{}_\nu (\Lambda^{}_{\mu\tau})
\cdot T^{}_l (\Lambda^{}_{\rm EW}) \right] \;
\label{eq:32}
%       (32)
\end{eqnarray}
in the safe $\tau$-flavor dominance approximation,
where $T^{}_l \simeq {\rm Diag}\{1 , 1 , 1 + \Delta_\tau\}$ and the
scale-dependent one-loop RGE evolution functions $I^{}_\nu$ and
$\Delta^{}_\tau$ are defined as~\cite{Zhang:2020lsd}
\begin{eqnarray}
I^{}_\nu (\mu) & \equiv & \exp\left[-\frac{1}{16 \pi^2}
\int^{\ln(\Lambda^{}_{\mu\tau}/\Lambda^{}_{\rm EW})}_{\ln(\mu/\Lambda^{}_{\rm EW})}
\alpha^{}_\nu(t) ~ {\rm d} t\right] \; ,
\nonumber \\
\Delta^{}_\tau (\mu) & \equiv & -\frac{C^{}_\nu}{16 \pi^2}
\int^{\ln(\Lambda^{}_{\mu\tau}/\Lambda^{}_{\rm EW})}_{\ln(\mu/\Lambda^{}_{\rm EW})}
y^2_\tau(t) ~ {\rm d} t \; .
\label{eq:33}
%       (33)
\end{eqnarray}
In Eq.~(\ref{eq:33}) $t\equiv \ln (\mu/\Lambda^{}_{\rm EW})$ with
$\Lambda^{}_{\rm EW} \leq \mu \leq \Lambda^{}_{\mu\tau}$,
$C^{}_\nu = -3/2$ and $\alpha^{}_\nu \simeq -3 g^2_2 + 6 y^2_t + \lambda$
in the SM framework with $g^{}_2$ being the gauge coupling constant of
weak interactions, $y^{}_t$ being the top-quark Yukawa coupling eigenvalue and
$\lambda$ being the Higgs self-coupling parameter; or $C^{}_\nu = 1$ and
$\alpha^{}_\nu \simeq -6 g^2_1/5 - 6 g^2_2 + 6 y^2_t$
in the minimal supersymmetric SM (MSSM) with $g^{}_1$ being the gauge
coupling constant of electromagnetic interactions. Moreover,
$y^{}_\tau = m^{}_\tau/\langle H\rangle$ in the SM or
$y^{}_\tau = m^{}_\tau/\left(\langle H\rangle \cos\beta\right)$ in the MSSM,
where $\cos\beta$ originates from $\tan\beta$ which is defined as the ratio
of the two Higgs doublets' vacuum expectation values. The smallness of
$y^{}_\tau$ means that $\Delta^{}_\tau$ is expected to be small and become
zero at $\mu = \Lambda^{}_{\mu\tau}$, as numerically illustrated in
Ref.~\cite{Zhang:2020lsd}. Therefore, we have
\begin{eqnarray}
M^{}_\nu (\Lambda^{}_{\rm EW}) \simeq M^{}_\nu (\Lambda^{}_{\mu\tau})
+ \left(\begin{matrix}
0 & 0 & \langle m\rangle^*_{e\mu} \cr
0 & 0 & \langle m\rangle^{}_{\mu\tau} \cr
\langle m\rangle^*_{e\mu} & ~\langle m\rangle^{}_{\mu\tau}~ &
2 \langle m\rangle^{}_{\tau\tau} \end{matrix} \right)
\Delta^{}_\tau (\Lambda^{}_{\rm EW}) \; ,
\label{eq:34}
%       (34)
\end{eqnarray}
where the second term with $\langle m\rangle^{}_{\alpha\beta}$ or its
complex conjugates at $\Lambda^{}_{\mu\tau}$ characterizes some small effects
of the $\mu$-$\tau$ reflection symmetry breaking. Note that
$M^{}_\nu (\Lambda^{}_{\rm EW})$ is of rank two as $M^{}_\nu (\Lambda^{}_{\mu\tau})$,
and thus it contains a vanishing eigenvalue.

Adopting the standard parametrization of the $3\times 3$ PMNS matrix $U$ in
Eq.~(\ref{eq:13}), we have the form invariance of $M^{}_\nu = U D^{}_\nu U^T$
with $D^{}_\nu \equiv {\rm Diag}\left\{m^{}_1, m^{}_2, m^{}_3\right\}$ at any
energy scale between $\Lambda^{}_{\rm EW}$ and $\Lambda^{}_{\mu\tau}$. A
combination of this decomposition and Eq.~(\ref{eq:34}) allows us to establish
some intriguing relations between the physical parameters of three active
Majorana neutrinos at these two scales. As for the neutrino masses, we explicitly
have
\begin{eqnarray}
m^{}_1 (\Lambda^{}_{\rm EW}) \hspace{-0.2cm} & \simeq & \hspace{-0.2cm} 0 \; ,
\nonumber \\
m^{}_2 (\Lambda^{}_{\rm EW}) \hspace{-0.2cm} & \simeq & \hspace{-0.2cm}
I^{}_\nu (\Lambda^{}_{\rm EW})
\Big[ 1 + \Delta^{}_\tau (\Lambda^{}_{\rm EW}) \left( 1 - s^2_{12} c^2_{13} \right)
\Big] m^{}_2 (\Lambda^{}_{\mu\tau}) \; ,
\nonumber \\
m^{}_3 (\Lambda^{}_{\rm EW}) \hspace{-0.2cm} & \simeq & \hspace{-0.2cm}
I^{}_\nu (\Lambda^{}_{\rm EW})
\Big[ 1 + \Delta^{}_\tau (\Lambda^{}_{\rm EW}) \hspace{0.05cm} c^2_{13} \Big]
m^{}_3 (\Lambda^{}_{\mu\tau}) \;
\label{eq:35}
%	    (35)
\end{eqnarray}
in the $m^{}_1 (\Lambda^{}_{\mu\tau}) = 0$ case; or
\begin{eqnarray}
m^{}_1 (\Lambda^{}_{\rm EW}) \hspace{-0.2cm} & \simeq & \hspace{-0.2cm}
I^{}_\nu (\Lambda^{}_{\rm EW})
\Big[ 1 + \Delta^{}_\tau (\Lambda^{}_{\rm EW}) \left( 1 - c^2_{12} c^2_{13} \right)
\Big] m^{}_1 (\Lambda^{}_{\mu\tau}) \; ,
\nonumber \\
m^{}_2 (\Lambda^{}_{\rm EW}) \hspace{-0.2cm} & \simeq & \hspace{-0.2cm}
I^{}_\nu (\Lambda^{}_{\rm EW})
\Big[ 1 + \Delta^{}_\tau (\Lambda^{}_{\rm EW}) \left( 1 - s^2_{12} c^2_{13} \right)
\Big] m^{}_2 (\Lambda^{}_{\mu\tau}) \; ,
\nonumber \\
m^{}_3 (\Lambda^{}_{\rm EW}) \hspace{-0.2cm} & \simeq & \hspace{-0.2cm} 0 \;
\label{eq:36}
%	    (36)
\end{eqnarray}
in the $m^{}_3 (\Lambda^{}_{\mu\tau}) = 0$ case, where
$\theta^{}_{23} (\Lambda^{}_{\mu\tau}) = \pi/4$,
$\delta (\Lambda^{}_{\mu\tau}) = \pm\pi/2$ and
$\sigma (\Lambda^{}_{\mu\tau}) = 0$ or $\pi/2$ have been input, and
$\theta^{}_{12}$ and $\theta^{}_{13}$ take their values measured at
$\Lambda^{}_{\rm EW}$. As for the flavor mixing angles and CP-violating
phases, let us define
\begin{eqnarray}
\Delta \theta^{}_{ij} \hspace{-0.2cm} & \equiv & \hspace{-0.2cm}
\theta^{}_{ij}(\Lambda_{\rm EW}^{})
- \theta^{}_{ij}(\Lambda_{\mu\tau}^{}) \; , \hspace{0.5cm}
\nonumber \\
\Delta \delta \hspace{-0.2cm} & \equiv & \hspace{-0.2cm}
\delta (\Lambda_{\rm EW}^{})
- \delta (\Lambda_{\mu\tau}^{}) \; ,
\nonumber \\
\Delta \sigma \hspace{-0.2cm} & \equiv & \hspace{-0.2cm}
\sigma (\Lambda_{\rm EW}^{})
-\sigma (\Lambda_{\mu\tau}^{}) \; ,
\label{eq:37}
%       (37)
\end{eqnarray}
to describe small deviations of their values at $\Lambda^{}_{\rm EW}$ from
those at $\Lambda^{}_{\mu\tau}$ where the translational $\mu$-$\tau$ reflection
flavor symmetry is exact. Following the generic analytical results obtained in
Ref.~\cite{Zhang:2020lsd}, we simply arrive at
\begin{eqnarray}
\Delta \theta^{}_{12} \hspace{-0.2cm} & \simeq & \hspace{-0.2cm}
-\frac{1}{2} c^{}_{12} s^{}_{12}
\left(1 - s^2_{13} \hspace{0.05cm} \zeta^{\eta^{}_\sigma}_{32} \right)
\Delta^{}_\tau (\Lambda^{}_{\rm EW}) \; ,
\nonumber \\
\Delta \theta^{}_{13} \hspace{-0.2cm} & \simeq & \hspace{-0.2cm}
-\frac{1}{2} c^{}_{13} s^{}_{13}
\left(c^2_{12} + s^2_{12} \hspace{0.05cm} \zeta^{\eta^{}_\sigma}_{32}\right)
\Delta^{}_\tau (\Lambda^{}_{\rm EW}) \; ,
\hspace{0.5cm}
\nonumber \\
\Delta \theta^{}_{23} \hspace{-0.2cm} & \simeq & \hspace{-0.2cm}
-\frac{1}{2} \left(s^2_{12} +
c^2_{12} \hspace{0.05cm} \zeta^{-\eta^{}_\sigma}_{32}\right)
\Delta^{}_\tau (\Lambda^{}_{\rm EW}) \;
\label{eq:38}
%	    (38)
\end{eqnarray}
in the $m^{}_1 (\Lambda^{}_{\mu\tau}) = 0$ case; or
\begin{eqnarray}
\Delta \theta^{}_{12} \hspace{-0.2cm} & \simeq & \hspace{-0.2cm}
-\frac{1}{2} c^{}_{12} s^{}_{12}
c^2_{13} \hspace{0.05cm} \zeta^{-\eta^{}_\sigma}_{21}
\Delta^{}_\tau (\Lambda^{}_{\rm EW}) \; , \hspace{0.5cm}
\nonumber \\
\Delta \theta^{}_{13} \hspace{-0.2cm} & \simeq & \hspace{-0.2cm}
+ \frac{1}{2} c^{}_{13} s^{}_{13}
\Delta^{}_\tau (\Lambda^{}_{\rm EW}) \; ,
\nonumber \\
\Delta \theta^{}_{23} \hspace{-0.2cm} & \simeq & \hspace{-0.2cm}
+\frac{1}{2} \Delta^{}_\tau (\Lambda^{}_{\rm EW}) \;
\label{eq:39}
%	    (39)
\end{eqnarray}
in the $m^{}_3 (\Lambda^{}_{\mu\tau}) = 0$ case, where
$\zeta^{}_{ij} \equiv \left(m^{}_i - m^{}_j\right)/\left(m^{}_i + m^{}_j\right)$
is defined at $\Lambda^{}_{\rm EW}$ and $\eta^{}_\sigma \equiv \cos 2\sigma = \pm 1$
denotes one of the two possible options of $\sigma (\Lambda^{}_{\mu\tau})$ in the
$\mu$-$\tau$ reflection symmetry limit~\cite{Zhou:2014sya}. Furthermore, we have
\begin{eqnarray}
\Delta \delta \hspace{-0.2cm} & \simeq & \hspace{-0.2cm}
-\frac{c^{}_{12} s^{}_{12} \eta^{}_\delta}{2 s^{}_{13}}
\left[1 - \zeta^{-\eta^{}_\sigma}_{32} + \frac{s^{2}_{13}}{c^{2}_{12} s^{2}_{12}}
\left(1 - s^4_{12} + c^4_{12} \hspace{0.05cm} \zeta^{-\eta^{}_\sigma}_{32}\right)\right]
\Delta^{}_\tau (\Lambda^{}_{\rm EW}) \; , \hspace{0.5cm}
\nonumber \\
\Delta \sigma \hspace{-0.2cm} & \simeq & \hspace{-0.2cm}
c^{}_{12} s^{}_{12} s^{}_{13} \eta^{}_\delta
\left(1 - \zeta^{-\eta^{}_\sigma}_{32}\right) \Delta^{}_\tau (\Lambda^{}_{\rm EW}) \;
\label{eq:40}
%	    (40)
\end{eqnarray}
in the $m^{}_1 (\Lambda^{}_{\mu\tau}) = 0$ case; or
\begin{eqnarray}
\Delta \delta \hspace{-0.2cm} & \simeq & \hspace{-0.2cm}
\frac{s^{}_{13} \eta^{}_\delta}{2 c^{}_{12} s^{}_{12}}
\hspace{0.05cm} \big(1 - 2 s^2_{12} - \zeta^{\eta^{}_\sigma}_{21}\big)
\hspace{0.05cm} \Delta^{}_\tau (\Lambda^{}_{\rm EW}) \; , \hspace{0.5cm}
\nonumber \\
\Delta \sigma \hspace{-0.2cm} & \simeq & \hspace{-0.2cm}
\frac{s^{}_{12} s^{}_{13} \eta^{}_\delta}{2 c^{}_{12}}
\hspace{0.05cm}\big(1 + \zeta^{\eta^{}_\sigma}_{21}\big) \hspace{0.05cm}
\Delta^{}_\tau (\Lambda^{}_{\rm EW}) \;
\label{eq:41}
%	    (41)
\end{eqnarray}
in the $m^{}_3 (\Lambda^{}_{\mu\tau}) = 0$ case,
where $\eta^{}_\delta \equiv \sin\delta = \pm 1$ stands for one of the two possible
options of $\delta (\Lambda^{}_{\mu\tau})$ in the $\mu$-$\tau$ reflection symmetry limit.
Some immediate comments are in order.
\begin{itemize}
\item     $m^{}_i (\Lambda^{}_{\mu\tau}) = 0$ will automatically lead to
$m^{}_i (\Lambda^{}_{\rm EW}) = 0$ (for $i = 1$ or $3$) as a natural consequence
of the one-loop RGE running. Note that a finite value of $m^{}_i (\Lambda^{}_{\rm EW})$
may result from $m^{}_i (\Lambda^{}_{\mu\tau}) = 0$ at the two-loop
level~\cite{Davidson:2006tg,Xing:2020ezi,Zhou:2021bqs}, but it is extremely tiny
(typically of ${\cal O}(10^{-13})~{\rm eV}$ in the SM or ${\cal O}(10^{-10})~{\rm eV}$
to ${\cal O}(10^{-8})~{\rm eV}$ in the MSSM) and can thus be treated as zero in
practice. Due to the smallness of $\Delta^{}_\tau$, the two nonzero neutrino masses
essentially have the same running effects characterized by $I^{}_\nu$.

\item     Among the three flavor mixing angles, $\theta^{}_{13}$ is most insensitive
to the RGE-induced quantum correction because of its smallness. That is to say,
$\Delta \theta^{}_{13}$ is proportional to $s^{}_{13}$ no matter whether
$m^{}_1 (\Lambda^{}_{\mu\tau}) = 0$ or $m^{}_3 (\Lambda^{}_{\mu\tau}) = 0$. The
smallness of $\Delta m^2_{21}$ implies that only $\Delta \theta^{}_{12}$ in the
$m^{}_3 (\Lambda^{}_{\mu\tau}) = 0$ case with $\sigma (\Lambda^{}_{\mu\tau}) = 0$
(i.e., $\eta^{}_\sigma = 1$) may be significantly enhanced. The deviation of
$\theta^{}_{23} (\Lambda^{}_{\rm EW})$ from $\pi/4$ is expected to be quite mild
unless $\tan\beta$ assumes a large value in the MSSM framework.

\item     One can see that the magnitude of $\Delta\sigma$ is suppressed by the
smallness of $\theta^{}_{13}$, no matter whether $m^{}_1 (\Lambda^{}_{\mu\tau}) = 0$
or $m^{}_3 (\Lambda^{}_{\mu\tau}) = 0$. In the latter case the magnitude of
$\Delta\delta$ is also suppressed by $s^{}_{13}$ but it can simultaneously
be enhanced by $\zeta^{\eta^{}_\sigma}_{21}$ with $\eta^{}_\sigma = -1$
(i.e., $\sigma (\Lambda^{}_{\mu\tau}) = \pi/2$), so can the magnitude of
$\Delta\sigma$. In comparison, the magnitude of $\Delta\delta$ is somewhat enhanced
by $1/s^{}_{13}$ in the $m^{}_1 (\Lambda^{}_{\mu\tau}) = 0$ case.
\end{itemize}
It is worth mentioning that the signs of $\Delta\theta^{}_{ij}$, $\Delta\delta$
and $\Delta\sigma$ in the SM and those in the MSSM are always opposite, because of
$\Delta^{}_\tau \propto C^{}_\nu$ having the opposite signs in these two frameworks.

To numerically illustrate our results obtained above, let us only consider the
possibility of $\delta = -\pi/2$ (i.e., $\eta^{}_\delta = -1$)
at $\Lambda^{}_{\mu\tau} \simeq 10^{14}~{\rm GeV}$
and input the best-fit values $s^{}_{12} \simeq 0.551$ and $s^{}_{13} \simeq 0.148$
at $\Lambda^{}_{\rm EW} \simeq 10^{2}~{\rm GeV}$. Using Eqs.~(\ref{eq:11}) and
(\ref{eq:12}), we obtain $\zeta^{}_{32} \simeq 0.707$ in the
$m^{}_1 (\Lambda^{}_{\mu\tau}) = 0$ case and $\zeta^{}_{21} \simeq 7.564
\times 10^{-3}$ in the $m^{}_3 (\Lambda^{}_{\mu\tau}) = 0$ case at
$\Lambda^{}_{\rm EW}$. As for the RGE-induced quantum corrections, we quote
$I^{}_\nu (\Lambda^{}_{\rm EW}) \simeq 0.748$ and
$\Delta^{}_\tau (\Lambda^{}_{\rm EW}) \simeq 2.822 \times 10^{-5}$ in the SM;
$I^{}_\nu (\Lambda^{}_{\rm EW}) \simeq 0.879$ and
$\Delta^{}_\tau (\Lambda^{}_{\rm EW}) \simeq -1.354 \times 10^{-3}$ in the MSSM
with $\tan\beta = 10$; or $I^{}_\nu (\Lambda^{}_{\rm EW}) \simeq 0.871$ and
$\Delta^{}_\tau (\Lambda^{}_{\rm EW}) \simeq -1.346 \times 10^{-2}$ in the MSSM
with $\tan\beta = 30$~\cite{Zhang:2020lsd}. Then we arrive at the numerical result
\begin{eqnarray}
\frac{m^{}_2 (\Lambda^{}_{\rm EW})}{m^{}_2 (\Lambda^{}_{\mu\tau})} \simeq
\frac{m^{}_3 (\Lambda^{}_{\rm EW})}{m^{}_3 (\Lambda^{}_{\mu\tau})}
\simeq I^{}_\nu (\Lambda^{}_{\rm EW}) \simeq
\left\{\begin{array}{l}
0.75 \quad\quad ({\rm SM}) \; ,
\\ \vspace{-0.6cm} \\
0.88 \quad\quad ({\rm MSSM}~{\rm with}~\tan\beta = 10) \; ,
\\ \vspace{-0.6cm} \\
0.87 \quad\quad ({\rm MSSM}~{\rm with}~\tan\beta = 30) \; ,
\end{array} \right.
\label{eq:42}
%	    (42)
\end{eqnarray}
in the $m^{}_1 (\Lambda^{}_{\mu\tau}) = 0$ case from Eq.~(\ref{eq:35});
and the same approximate result for the ratios
$m^{}_1 (\Lambda^{}_{\rm EW})/m^{}_1 (\Lambda^{}_{\mu\tau})$ and
$m^{}_2 (\Lambda^{}_{\rm EW})/m^{}_2 (\Lambda^{}_{\mu\tau})$ in the
$m^{}_3 (\Lambda^{}_{\mu\tau}) = 0$ case can be obtained from Eq.~(\ref{eq:36}).
The numerical values of $\Delta\theta^{}_{ij}$ (for $ij = 12, 13, 23$),
$\Delta\delta$ and $\Delta\sigma$ are calculated by using
Eqs.~(\ref{eq:38}) --- (\ref{eq:41}) for either $\sigma (\Lambda^{}_{\mu\tau}) = 0$
or $\sigma (\Lambda^{}_{\mu\tau}) = \pi/2$, and they are concisely listed in
Table~\ref{table:1}. These results are certainly compatible
with our above observations based on the analytical approximations. Two particular
remarks are in order.
%%%%%%%%%%%%%%%%%%%%%%%%%% Table 1 %%%%%%%%%%%%%%%%%%%%%%%%%%%%%%
%%%%%%%%%%%%%%%%%%%%%%%%%%%%%%%%%%%%%%%%%%%%%%%%%%%%%%%%%%%%%%%%%
\begin{table}[t]
\begin{center}
\caption{A numerical illustration of the RGE-induced quantum corrections to three
flavor mixing angles and two CP-violating phases in the canonical seesaw
model constrained by the translational $\mu$-$\tau$ reflection symmetry, where
$\Lambda^{}_{\mu\tau} \simeq 10^{14}~{\rm GeV}$ and
$\delta (\Lambda^{}_{\mu\tau}) = -\pi/2$ have been taken, and the results
of ${\cal O}(10^{-5})$ degrees or smaller have been denoted as $\sim 0^\circ$.}
\vspace{0.3cm}
\label{table:1}
\begin{tabular}{c|cc|cc|cc} \hline\hline
~$m^{}_1 (\Lambda^{}_{\mu\tau}) = 0$~
& \multicolumn{2}{c}{SM}
& \multicolumn{2}{c}{MSSM ($\tan\beta = 10$)}
& \multicolumn{2}{c}{MSSM ($\tan\beta = 30$)}
\\ \hline
$\sigma (\Lambda^{}_{\mu\tau})$
& $0$
& $\pi/2$
& $0$
& $\pi/2$
& $0$
& $\pi/2$
\\ \hline
%------------------------------------------------------------
$\Delta\theta^{}_{12}$
& $\sim 0^\circ$
& $\sim 0^\circ$
& $0.018^\circ$
& $0.017^\circ$
& $0.175^\circ$
& $0.172^\circ$
\\
$\Delta\theta^{}_{13}$
& $\sim 0^\circ$
& $\sim 0^\circ$
& $0.005^\circ$
& $0.006^\circ$
& $0.051^\circ$
& $0.064^\circ$
\\
$\Delta\theta^{}_{23}$
& $-0.001^\circ$
& $-0.0006^\circ$
& $0.050^\circ$
& $0.031^\circ$
& $0.497^\circ$
& $0.307^\circ$
\\
$\Delta\delta$
& $-0.0004^\circ$
& $0.001^\circ$
& $0.021^\circ$
& $-0.056^\circ$
& $0.207^\circ$
& $-0.552^\circ$
\\
$\Delta\sigma$
& $\sim 0^\circ$
& $\sim 0^\circ$
& $-0.002^\circ$
& $0.0015^\circ$
& $-0.022^\circ$
& $0.015^\circ$
\\ \hline\hline
%============================================================
~~$m^{}_3 (\Lambda^{}_{\mu\tau}) = 0$~~
& \multicolumn{2}{c}{SM}
& \multicolumn{2}{c}{MSSM ($\tan\beta = 10$)}
& \multicolumn{2}{c}{MSSM ($\tan\beta = 30$)}
\\ \hline
$\sigma (\Lambda^{}_{\mu\tau})$
& $0$
& $\pi/2$
& $0$
& $\pi/2$
& $0$
& $\pi/2$
\\ \hline
%------------------------------------------------------------
$\Delta\theta^{}_{12}$
& $-0.048^\circ$
& $\sim 0^\circ$
& $2.306^\circ$
& $0.0001^\circ$
& $22.925^\circ$
& $0.001^\circ$
\\
$\Delta\theta^{}_{13}$
& $0.0001^\circ$
& $0.0001^\circ$
& $-0.006^\circ$
& $-0.006^\circ$
& $-0.056^\circ$
& $-0.056^\circ$
\\
$\Delta\theta^{}_{23}$
& $0.0008^\circ$
& $0.0008^\circ$
& $-0.039^\circ$
& $-0.039^\circ$
& $-0.386^\circ$
& $-0.386^\circ$
\\
$\Delta\delta$
& $-0.0001^\circ$
& $0.034^\circ$
& $0.005^\circ$
& $-1.646^\circ$
& $0.048^\circ$
& $-16.360^\circ$
\\
$\Delta\sigma$
& $-0.00008^\circ$
& $-0.011^\circ$
& $0.004^\circ$
& $0.505^\circ$
& $0.038^\circ$
& $5.019^\circ$
%------------------------------------------------------------
\\ \hline\hline
%%%%%%%%%%%%%%%%%%%%%%%%%%%%%%%%%%%%%%%%%%%%%%%%%%%%%%%%%%%%%
\end{tabular}
\end{center}
\end{table}
%%%%%%%%%%%%%%%%%%%%%%%%%%%%%%%%%%%%%%%%%%%%%%%%%%%%%%%%%%%%%%%%%
%%%%%%%%%%%%%%%%%%%%%%%%%%%%%%%%%%%%%%%%%%%%%%%%%%%%%%%%%%%%%%%%%
\begin{itemize}
\item     Now that the neutrino mass spectrum has been fixed by the
translational flavor symmetry (i.e., either $m^{}_1 = 0$ or
$m^{}_3 = 0$ at $\Lambda^{}_{\mu\tau}$) and the observed neutrino mass-squared
differences, there is no freedom to adjust the strength of the RGE-induced
quantum corrections in the SM case. That is why the numerical results of
$\Delta\theta^{}_{ij}$, $\Delta\delta$ and $\Delta\sigma$ are all too small to be
experimentally distinguishable. In the MSSM case taking relatively large values of
$\tan\beta$ may help enhance the significance of the RGE-induced effects, simply
because $\Delta{}_\tau (\Lambda^{}_{\rm EW}) \propto \left(1 + \tan^2\beta\right)$
holds.

\item     Table~\ref{table:1} shows that $\Delta\theta^{}_{12}$ can be significantly
enhanced for $m^{}_3 (\Lambda^{}_{\mu\tau}) = 0$ and $\sigma (\Lambda^{}_{\mu\tau}) = 0$
in the MSSM framework with a sufficiently large input of $\tan\beta$. The reason is
simply that $\Delta\theta^{}_{12}$ is proportional to both
$\zeta^{-1}_{21} = \left(m^{}_1 + m^{}_2\right)^2/\Delta m^2_{21} \simeq 132.2$
and $\left(1 + \tan^2\beta\right) \gtrsim 10^2$ for $\tan\beta \gtrsim 10$
and thus gets enhanced in this case. In comparison, $\Delta\delta$ and $\Delta\sigma$
can be remarkably enhanced for the same reasons if $m^{}_3 (\Lambda^{}_{\mu\tau}) = 0$
and $\sigma (\Lambda^{}_{\mu\tau}) = \pi/2$ are taken in the MSSM, although these two
quantities are apparently suppressed by the smallness of $\theta^{}_{13}$ as shown
in Eq.~(\ref{eq:41}).
\end{itemize}
In addition, the small magnitude of $\Delta\theta^{}_{23}$ in the MSSM with
$\tan\beta = 30$ implies that the RGE-induced quantum correction to
$\theta^{}_{23} (\Lambda^{}_{\mu\tau}) = \pi/4$ at $\Lambda^{}_{\rm EW}$ is in
general too mild to resolve the octant issue of $\theta^{}_{23}$ in the minimal
seesaw scenario.

If the values of $\theta^{}_{23}$ and $\delta$ to be
measured in the future long-baseline neutrino oscillation experiments (e.g.,
Hyper-Kamiokande~\cite{Hyper-KamiokandeProto-:2015xww} and DUNE~\cite{DUNE:2015lol})
turn out to be remarkably different from their values constrained by the
translational $\mu$-$\tau$ reflection symmetry under discussion, it will be
difficult to naturally account for such a discrepancy with the help of the
above RGE-induced soft symmetry breaking effects. In this case one may either
invoke an {\it explicit} $\mu$-$\tau$ reflection symmetry breaking scenario to
fit the experimental data (see, e.g., a recent scenario proposed in
Ref.~\cite{Fukuyama:2020swd}), which certainly seems contrived, or go beyond the
$\mu$-$\tau$ reflection symmetry by invoking a larger flavor symmetry group for
model building. In either case it will be useful to explore the direct correlation
between thermal leptogenesis and CP violation in neutrino oscillations (see, e.g.,
the pioneering attempts in Refs.~\cite{Frampton:2002qc,Endoh:2002wm} and a
recent comprehensive review in Ref.\cite{Xing:2020ald}).

\section{Summary}

The strong hierarchies associated with the mass spectra of charged leptons,
up-type quarks and down-type quarks, together with a clear hierarchy of
the off-diagonal CKM matrix elements, strongly indicate that the flavors
of charged fermions should have specific structures instead of a random
nature. This expectation has been further strengthened in recent years with
the very observation that the PMNS lepton flavor mixing matrix $U$ exhibits an
approximate $\mu$-$\tau$ interchange symmetry, although whether the three
active neutrinos have a normal, inverted or nearly degenerate mass spectrum
is still an open question. So far a lot of phenomenological efforts have been
made towards deeper understanding of the issues of tiny neutrino masses and
significant lepton flavor mixing and CP violation with the help of current
neutrino oscillation data, but new ideas are always called for because we
still have a long way to go before the true theory finally emerges.

Along this line of thought, we have coherently combined two simple but
suggestive flavor symmetries --- the $\mu$-$\tau$ reflection symmetry and the
translational flavor symmetry by proposing a {\it translational $\mu$-$\tau$
reflection symmetry} for the effective Majorana neutrino mass term
${\cal L}^{}_{\rm mass}$ in this work. Namely, ${\cal L}^{}_{\rm mass}$
is required to keep invariant under the transformations
$\nu^{}_{e \rm L} \to (\nu^{}_{e \rm L})^c + U^*_{e i} z^{c}_\nu$,
$\nu^{}_{\mu \rm L} \to (\nu^{}_{\tau \rm L})^c + U^*_{\tau i} z^{c}_\nu$ and
$\nu^{}_{\tau \rm L} \to (\nu^{}_{\mu \rm L})^c + U^*_{\mu i} z^{c}_\nu$,
where $z^{}_\nu$ is a constant spinor field and $U^{}_{\alpha i}$ are the
PMNS matrix elements corresponding to $m^{}_i = 0$ (for $\alpha = e, \mu, \tau$
and $i = 1$ or $3$). Extending such a flavor symmetry to the canonical seesaw
mechanism by requiring the relevant neutrino mass terms to be
invariant under similar translational $\mu$-$\tau$ reflection
transformations for the right-handed neutrino fields, we show that a
minimal seesaw scenario can naturally be reproduced and it automatically
respects the $\mu$-$\tau$ reflection symmetry. As a by-product, the soft
breaking effects of this flavor symmetry have been studied with the help
of the one-loop RGEs.

We admit that the translational $\mu$-$\tau$ reflection symmetry is not
a real flavor symmetry for the whole Lagrangian of a Majorana neutrino mass
model. Instead, it is just an effective or working flavor symmetry for the
effective neutrino mass term, but it can greatly help constrain the corresponding
flavor textures. Motivated by the principle of {\it Occam's razor}, we find
that such simple but viable and predictive neutrino mass models really
deserve attention. First, their predictions can be more easily tested in the near
future. Second, it is very likely that the empirical flavor symmetries of this kind
are actually the residual symmetries of some larger flavor symmetry groups in an
underlying fundamental and UV-complete flavor theory. In this sense,
we may argue that the translational $\mu$-$\tau$
reflection symmetry under discussion, among many other existing flavor
symmetries, might be located in the {\it neutrino landscape} instead of the
{\it neutrino swampland}~\cite{Vafa:2005ui,Palti:2019pca,Xing:2021SUSY}.

\vspace{0.3cm}
\begin{flushleft}
{\large\bf Acknowledgments}
\end{flushleft}

I am grateful to Dr. Di Zhang for very useful discussions
and for his help in correcting a fatal ``full stop" error
in my LaTeX file. This work is supported in part by the National
Natural Science Foundation of China under grants No. 12075254 and No. 11835013.

%\newpage


\begin{thebibliography}{99}

\bibitem{Zyla:2020zbs}
P.~A.~Zyla \textit{et al.} [Particle Data Group],
``Review of Particle Physics,''
PTEP \textbf{2020} (2020) no.8, 083C01.

\bibitem{Pontecorvo:1957cp}
B.~Pontecorvo,
``Mesonium and anti-mesonium,''
Sov.\ Phys.\ JETP {\bf 6} (1957) 429
[Zh.\ Eksp.\ Teor.\ Fiz.\  {\bf 33} (1957) 549].

\bibitem{Maki:1962mu}
Z.~Maki, M.~Nakagawa and S.~Sakata,
``Remarks on the unified model of elementary particles,''
Prog.\ Theor.\ Phys.\  {\bf 28} (1962) 870.

\bibitem{Pontecorvo:1967fh}
B.~Pontecorvo,
``Neutrino experiments and the problem of conservation of leptonic charge,''
Sov.\ Phys.\ JETP {\bf 26} (1968) 984
[Zh.\ Eksp.\ Teor.\ Fiz.\  {\bf 53} (1967) 1717].

\bibitem{Capozzi:2021fjo}
F.~Capozzi, E.~Di Valentino, E.~Lisi, A.~Marrone, A.~Melchiorri and A.~Palazzo,
``Unfinished fabric of the three neutrino paradigm,''
Phys. Rev. D \textbf{104} (2021) no.8, 083031
[arXiv:2107.00532 [hep-ph]].

\bibitem{Gonzalez-Garcia:2021dve}
M.~C.~Gonzalez-Garcia, M.~Maltoni and T.~Schwetz,
``NuFIT: Three-flavour global analyses of neutrino oscillation experiments,''
Universe \textbf{7} (2021) no.12, 459
[arXiv:2111.03086 [hep-ph]].

\bibitem{Cabibbo:1963yz}
  N.~Cabibbo,
  Unitary Symmetry and Leptonic Decays,
  Phys.\ Rev.\ Lett.\  {\bf 10} (1963) 531.

\bibitem{Kobayashi:1973fv}
  M.~Kobayashi and T.~Maskawa,
  CP Violation in the Renormalizable Theory of Weak Interaction,
  Prog.\ Theor.\ Phys.\  {\bf 49} (1973) 652.

\bibitem{T2K:2019bcf}
K.~Abe \textit{et al.} [T2K],
``Constraint on the matter\textendash{}antimatter symmetry-violating phase in neutrino
oscillations,''
Nature \textbf{580} (2020) no.7803, 339-344
[erratum: Nature \textbf{583} (2020) no.7814, E16]
[arXiv:1910.03887 [hep-ex]].

\bibitem{NOvA:2019cyt}
M.~A.~Acero \textit{et al.} [NOvA],
``First Measurement of Neutrino Oscillation Parameters using Neutrinos and Antineutrinos by NOvA,''
Phys. Rev. Lett. \textbf{123} (2019) no.15, 151803
[arXiv:1906.04907 [hep-ex]].

\bibitem{Altarelli:2010gt}
G.~Altarelli and F.~Feruglio,
``Discrete flavor symmetries and models of neutrino mixing,''
Rev. Mod. Phys. \textbf{82} (2010), 2701-2729
[arXiv:1002.0211 [hep-ph]].

\bibitem{Ishimori:2010au}
H.~Ishimori, T.~Kobayashi, H.~Ohki, Y.~Shimizu, H.~Okada and M.~Tanimoto,
``Non-Abelian discrete symmetries in particle physics,''
Prog. Theor. Phys. Suppl. \textbf{183} (2010), 1-163
[arXiv:1003.3552 [hep-th]].

\bibitem{King:2013eh}
S.~F.~King and C.~Luhn,
``Neutrino mass and mixing with discrete symmetry,''
Rept. Prog. Phys. \textbf{76} (2013), 056201
[arXiv:1301.1340 [hep-ph]].

\bibitem{Xing:2015fdg}
Z.~z.~Xing and Z.~h.~Zhao,
``A review of \ensuremath{\mu}-\ensuremath{\tau} flavor symmetry in neutrino physics,''
Rept. Prog. Phys. \textbf{79} (2016) no.7, 076201
[arXiv:1512.04207 [hep-ph]].

\bibitem{Xing:2019vks}
Z.~z.~Xing,
``Flavor structures of charged fermions and massive neutrinos,''
Phys. Rept. \textbf{854} (2020), 1-147
[arXiv:1909.09610 [hep-ph]].

\bibitem{Feruglio:2019ybq}
F.~Feruglio and A.~Romanino,
``Lepton flavor symmetries,''
Rev. Mod. Phys. \textbf{93} (2021) no.1, 015007
[arXiv:1912.06028 [hep-ph]].

\bibitem{Xing:2022uax}
Z.~z.~Xing,
``The $\mu$-$\tau$ reflection symmetry of Majorana neutrinos,''
[arXiv:2210.11922 [hep-ph]].

\bibitem{Harrison:2002et}
P.~F.~Harrison and W.~G.~Scott,
``$\mu$-$\tau$ reflection symmetry in lepton mixing and neutrino oscillations,''
Phys. Lett. B \textbf{547} (2002), 219-228
[arXiv:hep-ph/0210197 [hep-ph]].

\bibitem{Kleppe:1995zz}
A.~Kleppe,
``Extending the standard model with two right-handed neutrinos,''
in {\it Neutrino physics} (Proceedings of the 3rd Tallinn Symposium, Lohusalu, Estonia,
October 8-11, 1995), page 118-125.

\bibitem{Ma:1998zg}
E.~Ma, D.~P.~Roy and U.~Sarkar,
``A Seesaw model for atmospheric and solar neutrino oscillations,''
Phys. Lett. B \textbf{444} (1998), 391-396
[arXiv:hep-ph/9810309 [hep-ph]].

\bibitem{Frampton:2002qc}
P.~H.~Frampton, S.~L.~Glashow and T.~Yanagida,
``Cosmological sign of neutrino CP violation,''
Phys. Lett. B \textbf{548} (2002), 119-121
[arXiv:hep-ph/0208157 [hep-ph]].

\bibitem{Friedberg:2006it}
R.~Friedberg and T.~D.~Lee,
``A possible relation between the neutrino mass matrix and the neutrino mapping matrix,''
HEPNP \textbf{30} (2006), 591-598
[arXiv:hep-ph/0606071 [hep-ph]].

\bibitem{Lee:2008zzh}
T.~D.~Lee,
``Symmetry and asymmetry,''
Nucl. Phys. A \textbf{805} (2008), 54-71

\bibitem{Friedberg:2007ba}
R.~Friedberg and T.~D.~Lee,
``Jarlskog invariant of the neutrino mapping matrix,''
Annals Phys. \textbf{323} (2008), 1677-1691
[arXiv:0709.1526 [hep-ph]].

\bibitem{Friedberg:2007uk}
R.~Friedberg and T.~D.~Lee,
``Hidden symmetry of the CKM and neutrino mapping matrices,''
Annals Phys. \textbf{323} (2008), 1087-1105
[arXiv:0705.4156 [hep-ph]].

\bibitem{Friedberg:2009fb}
R.~Friedberg and T.~D.~Lee,
``A timeon model of quark and lepton mass matrices,''
Annals Phys. \textbf{324} (2009), 2196-2225
[arXiv:0904.1640 [hep-ph]].

\bibitem{Friedberg:2010zt}
R.~Friedberg and T.~D.~Lee,
``Deviations of the lepton mapping matrix from the Harrison-Perkins-Scott form,''
Chin. Phys. C \textbf{34} (2010), 1547-1555
[erratum: Chin. Phys. C \textbf{34} (2010), 1905]
[arXiv:1008.0453 [hep-ph]].

\bibitem{Xing:2006ms}
Z.~z.~Xing and S.~Zhou,
``Tri-bimaximal neutrino mixing and flavor-dependent resonant leptogenesis,''
Phys. Lett. B \textbf{653} (2007), 278-287
[arXiv:hep-ph/0607302 [hep-ph]].

\bibitem{King:2015dvf}
S.~F.~King,
``Littlest seesaw,''
JHEP \textbf{02} (2016), 085
[arXiv:1512.07531 [hep-ph]].

\bibitem{King:2016yef}
S.~F.~King, J.~Zhang and S.~Zhou,
``Renormalisation group corrections to the littlest seesaw model and maximal
atmospheric mixing,''
JHEP \textbf{12} (2016), 023
[arXiv:1609.09402 [hep-ph]].

\bibitem{Liu:2017frs}
Z.~C.~Liu, C.~X.~Yue and Z.~h.~Zhao,
``Neutrino $\mu$-$\tau$ reflection symmetry and its breaking in the minimal seesaw,''
JHEP \textbf{10} (2017), 102
[arXiv:1707.05535 [hep-ph]].

\bibitem{Nath:2018hjx}
N.~Nath, Z.~z.~Xing and J.~Zhang,
``$\mu$-$\tau$ reflection symmetry embedded in minimal seesaw,''
Eur. Phys. J. C \textbf{78} (2018) no.4, 289
[arXiv:1801.09931 [hep-ph]].

\bibitem{Samanta:2017kce}
R.~Samanta, P.~Roy and A.~Ghosal,
``Consequences of minimal seesaw with complex $\mu\tau$ antisymmetry of neutrinos,''
JHEP \textbf{06} (2018), 085
[arXiv:1712.06555 [hep-ph]].

\bibitem{King:2018kka}
S.~F.~King and C.~C.~Nishi,
``$\mu$-$\tau$ symmetry and the Littlest Seesaw,''
Phys. Lett. B \textbf{785} (2018), 391-398
[arXiv:1807.00023 [hep-ph]].

\bibitem{Nath:2018xih}
N.~Nath,
``$\mu$-$\tau$ reflection symmetry and its explicit breaking for leptogenesis in a
minimal seesaw model,''
Mod. Phys. Lett. A \textbf{34} (2019) no.39, 1950329
[arXiv:1808.05062 [hep-ph]].

\bibitem{King:2019tbt}
S.~F.~King and Y.~L.~Zhou,
``Littlest $\mu$-$\tau$ seesaw,''
JHEP \textbf{05} (2019), 217
[arXiv:1901.06877 [hep-ph]].

\bibitem{Xing:2020ald}
Z.~z.~Xing and Z.~h.~Zhao,
``The minimal seesaw and leptogenesis models,''
Rept. Prog. Phys. \textbf{84} (2021) no.6, 066201
[arXiv:2008.12090 [hep-ph]].

\bibitem{Kang:2021stv}
S.~K.~Kang,
``Low-energy CP violation and leptogenesis in a minimal seesaw model,''
J. Korean Phys. Soc. \textbf{78} (2021) no.9, 743-749

\bibitem{Zhao:2020bzx}
Z.~h.~Zhao,
``Renormalization group evolution induced leptogenesis in the minimal seesaw model
with the trimaximal mixing and $\mu$-$\tau$ reflection symmetry,''
JHEP \textbf{11} (2021), 170
[arXiv:2003.00654 [hep-ph]].

\bibitem{Zhao:2021dwc}
Z.~h.~Zhao,
``A combination of the neutrino trimaximal mixings and $\mu$-$\tau$ reflection symmetry
in the type-I seesaw model,''
Eur. Phys. J. C \textbf{82} (2022) no.5, 436
[arXiv:2111.12639 [hep-ph]].

\bibitem{Xing:2006xa}
Z.~z.~Xing, H.~Zhang and S.~Zhou,
``Nearly tri-bimaximal neutrino mixing and CP violation from $\mu$-$\tau$ symmetry breaking,''
Phys. Lett. B \textbf{641} (2006), 189-197
[arXiv:hep-ph/0607091 [hep-ph]].

\bibitem{Huang:2008ri}
C.~S.~Huang, T.~j.~Li, W.~Liao and S.~H.~Zhu,
``Generalization of Friedberg-Lee Symmetry,''
Phys. Rev. D \textbf{78} (2008), 013005
[arXiv:0803.4124 [hep-ph]].

\bibitem{Jarlskog:2007qy}
C.~Jarlskog,
``Neutrino sector with Majorana mass terms and Friedberg-Lee symmetry,''
Phys. Rev. D \textbf{77} (2008), 073002
[arXiv:0712.0903 [hep-ph]].

\bibitem{He:2009pt}
X.~G.~He and W.~Liao,
``The Friedberg-Lee Symmetry and Minimal Seesaw Model,''
Phys. Lett. B \textbf{681} (2009), 253-256
[arXiv:0909.1463 [hep-ph]].

\bibitem{Razzaghi:2012rr}
N.~Razzaghi and S.~S.~Gousheh,
``Generalized Friedberg-Lee model for CP violation in neutrino physics,''
Phys. Rev. D \textbf{86} (2012), 053006
doi:10.1103/PhysRevD.86.053006
[arXiv:1211.4389 [hep-ph]].

\bibitem{Zhao:2015bza}
Z.~h.~Zhao,
``Modified Friedberg-Lee symmetry for neutrino mixing,''
Phys. Rev. D \textbf{92} (2015) no.11, 113001
[arXiv:1509.06915 [hep-ph]].

\bibitem{Xing:2021zxf}
Z.~z.~Xing,
``A translational flavor symmetry in the mass terms of Dirac and Majorana fermions,''
J. Phys. G \textbf{49} (2022) no.2, 025003
[arXiv:2102.03050 [hep-ph]].

\bibitem{Volkov:1973ix}
D.~V.~Volkov and V.~P.~Akulov,
``Is the neutrino a Goldstone particle?,''
Phys. Lett. B \textbf{46} (1973), 109-110

\bibitem{deWit:1975xci}
B.~de Wit and D.~Z.~Freedman,
``Phenomenology of Goldstone neutrinos,''
Phys. Rev. Lett. \textbf{35} (1975), 827

\bibitem{Araki:2009grl}
T.~Araki and R.~Takahashi,
``Tri-Bimaximal Mixing from Twisted Friedberg-Lee Symmetry,''
Eur. Phys. J. C \textbf{63} (2009), 521-526
[arXiv:0811.0905 [hep-ph]].

\bibitem{Araki:2009kp}
T.~Araki and C.~Q.~Geng,
``Leptogenesis in model with Friedberg-Lee symmetry,''
Phys. Lett. B \textbf{680} (2009), 343-350
[arXiv:0906.1903 [hep-ph]].

\bibitem{Sinha:2018uop}
R.~Sinha, S.~Bhattacharya and R.~Samanta,
``Phenomenological implications of the Friedberg-Lee transformation in a neutrino
mass model with $\mu\tau$-flavored CP symmetry,''
JHEP \textbf{03} (2019), 081
[arXiv:1810.05391 [hep-ph]].

\bibitem{Fritzsch:1975sr}
H.~Fritzsch, M.~Gell-Mann and P.~Minkowski,
``Vector-like weak Currents and new elementary fermions,''
Phys. Lett. B \textbf{59} (1975), 256-260

\bibitem{Minkowski:1977sc}
  P.~Minkowski,
  ``$\mu \to e\gamma$ at a rate of one out of $10^{9}$ muon decays?,''
  Phys.\ Lett.\  {\bf 67B} (1977) 421.

\bibitem{Yanagida:1979as}
  T.~Yanagida,
  ``Horizontal gauge symmetry and masses of neutrinos,''
  Conf.\ Proc.\ C {\bf 7902131} (1979) 95.

\bibitem{GellMann:1980vs}
  M.~Gell-Mann, P.~Ramond and R.~Slansky,
  ``Complex spinors and unified theories,''
  Conf.\ Proc.\ C {\bf 790927} (1979) 315
  [arXiv:1306.4669 [hep-th]].

\bibitem{Glashow:1979nm}
  S.~L.~Glashow,
  ``The future of elementary particle physics,''
  NATO Sci.\ Ser.\ B {\bf 61} (1980) 687.

\bibitem{Mohapatra:1979ia}
  R.~N.~Mohapatra and G.~Senjanovic,
  ``Neutrino mass and spontaneous parity nonconservation,''
  Phys.\ Rev.\ Lett.\  {\bf 44} (1980) 912.

\bibitem{Rodejohann:2011mu}
W.~Rodejohann,
``Neutrino-less Double Beta Decay and Particle Physics,''
Int. J. Mod. Phys. E \textbf{20} (2011), 1833-1930
[arXiv:1106.1334 [hep-ph]].

\bibitem{Bilenky:2014uka}
S.~M.~Bilenky and C.~Giunti,
``Neutrinoless Double-Beta Decay: a Probe of Physics Beyond the Standard Model,''
Int. J. Mod. Phys. A \textbf{30} (2015) no.04n05, 1530001
[arXiv:1411.4791 [hep-ph]].

\bibitem{Dolinski:2019nrj}
M.~J.~Dolinski, A.~W.~P.~Poon and W.~Rodejohann,
``Neutrinoless Double-Beta Decay: Status and Prospects,''
Ann. Rev. Nucl. Part. Sci. \textbf{69} (2019), 219-251
[arXiv:1902.04097 [nucl-ex]].

\bibitem{LHCb:2018roe}
R.~Aaij \textit{et al.} [LHCb],
``Physics case for an LHCb Upgrade II --- Opportunities in flavour physics, and
beyond, in the HL-LHC era,''
[arXiv:1808.08865 [hep-ex]].

\bibitem{Belle-II:2018jsg}
E.~Kou \textit{et al.} [Belle-II],
``The Belle II Physics Book,''
PTEP \textbf{2019} (2019) no.12, 123C01
[erratum: PTEP \textbf{2020} (2020) no.2, 029201]
[arXiv:1808.10567 [hep-ex]].

\bibitem{Xing:2017mkx}
Z.~z.~Xing, D.~Zhang and J.~y.~Zhu,
``The $\mu$-$\tau$ reflection symmetry of Dirac neutrinos and its breaking effect via
quantum corrections,''
JHEP \textbf{11} (2017), 135
[arXiv:1708.09144 [hep-ph]].

\bibitem{Xing:2019edp}
Z.~Z.~Xing and D.~Zhang,
``Seesaw mirroring between light and heavy Majorana neutrinos with the help of the
S$_{3}$ reflection symmetry,''
JHEP \textbf{03} (2019), 184
[arXiv:1901.07912 [hep-ph]].

\bibitem{Weinberg:1979sa}
S.~Weinberg,
``Baryon and Lepton Nonconserving Processes,''
Phys. Rev. Lett. \textbf{43} (1979), 1566-1570

\bibitem{Xing:2007uq}
Z.~z.~Xing,
``Massive and Massless Neutrinos on Unbalanced Seesaws,''
Chin. Phys. C \textbf{32} (2008), 96-99
[arXiv:0706.0052 [hep-ph]].

\bibitem{StueckelbergdeBreidenbach:1952pwl}
E.~C.~G.~Stueckelberg de Breidenbach and A.~Petermann,
``Normalization of constants in the quanta theory,''
Helv. Phys. Acta \textbf{26} (1953), 499-520

\bibitem{Gell-Mann:1954yli}
M.~Gell-Mann and F.~E.~Low,
``Quantum electrodynamics at small distances,''
Phys. Rev. \textbf{95} (1954), 1300-1312

\bibitem{Bogolyubov:1956gh}
N.~N.~Bogolyubov and D.~V.~Shirkov,
``Charge renormalization group in quantum field theory,''
Nuovo Cim. \textbf{3} (1956), 845-863

\bibitem{Wilson:1971bg}
K.~G.~Wilson,
``Renormalization group and critical phenomena. 1. Renormalization group and the Kadanoff
scaling picture,''
Phys. Rev. B \textbf{4} (1971), 3174-3183

\bibitem{Wilson:1971dh}
K.~G.~Wilson,
``Renormalization group and critical phenomena. 2. Phase space cell analysis of critical
behavior,''
Phys. Rev. B \textbf{4} (1971), 3184-3205

\bibitem{Gross:1973id}
D.~J.~Gross and F.~Wilczek,
``Ultraviolet Behavior of Nonabelian Gauge Theories,''
Phys. Rev. Lett. \textbf{30} (1973), 1343-1346

\bibitem{Politzer:1973fx}
H.~D.~Politzer,
``Reliable Perturbative Results for Strong Interactions?,''
Phys. Rev. Lett. \textbf{30} (1973), 1346-1349

\bibitem{Ohlsson:2013xva}
T.~Ohlsson and S.~Zhou,
``Renormalization group running of neutrino parameters,''
Nature Commun. \textbf{5} (2014), 5153
[arXiv:1311.3846 [hep-ph]].

\bibitem{Chankowski:1993tx}
P.~H.~Chankowski and Z.~Pluciennik,
``Renormalization group equations for seesaw neutrino masses,''
Phys. Lett. B \textbf{316} (1993), 312-317
[arXiv:hep-ph/9306333 [hep-ph]].

\bibitem{Babu:1993qv}
K.~S.~Babu, C.~N.~Leung and J.~T.~Pantaleone,
``Renormalization of the neutrino mass operator,''
Phys. Lett. B \textbf{319} (1993), 191-198
[arXiv:hep-ph/9309223 [hep-ph]].

\bibitem{Casas:1999tg}
J.~A.~Casas, J.~R.~Espinosa, A.~Ibarra and I.~Navarro,
``General RG equations for physical neutrino parameters and their phenomenological
implications,''
Nucl. Phys. B \textbf{573} (2000), 652-684
[arXiv:hep-ph/9910420 [hep-ph]].

\bibitem{Antusch:2001ck}
S.~Antusch, M.~Drees, J.~Kersten, M.~Lindner and M.~Ratz,
``Neutrino mass operator renormalization revisited,''
Phys. Lett. B \textbf{519} (2001), 238-242
[arXiv:hep-ph/0108005 [hep-ph]].

\bibitem{Ellis:1999my}
J.~R.~Ellis and S.~Lola,
``Can neutrinos be degenerate in mass?,''
Phys. Lett. B \textbf{458} (1999), 310-321
[arXiv:hep-ph/9904279 [hep-ph]].

\bibitem{Fritzsch:1999ee}
H.~Fritzsch and Z.~z.~Xing,
``Mass and flavor mixing schemes of quarks and leptons,''
Prog. Part. Nucl. Phys. \textbf{45} (2000), 1-81
[arXiv:hep-ph/9912358 [hep-ph]].

\bibitem{Mei:2003gn}
J.~w.~Mei and Z.~z.~Xing,
``Radiative corrections to neutrino mixing and CP violation in the minimal seesaw
model with leptogenesis,''
Phys. Rev. D \textbf{69} (2004), 073003
[arXiv:hep-ph/0312167 [hep-ph]].

\bibitem{Zhang:2020lsd}
D.~Zhang,
``Integral solutions to the one-loop renormalization-group equations for lepton
flavor mixing parameters and the Jarlskog invariant,''
Nucl. Phys. B \textbf{961} (2020), 115260
[arXiv:2007.12976 [hep-ph]].

\bibitem{Zhou:2014sya}
Y.~L.~Zhou,
``$\mu$-$\tau$ reflection symmetry and radiative corrections,''
[arXiv:1409.8600 [hep-ph]].

\bibitem{Davidson:2006tg}
S.~Davidson, G.~Isidori and A.~Strumia,
``The Smallest neutrino mass,''
Phys. Lett. B \textbf{646} (2007), 100-104
[arXiv:hep-ph/0611389 [hep-ph]].

\bibitem{Xing:2020ezi}
Z.~z.~Xing and D.~Zhang,
``On the two-loop radiative origin of the smallest neutrino mass and the associated
Majorana CP phase,''
Phys. Lett. B \textbf{807} (2020), 135598
[arXiv:2005.05171 [hep-ph]].

\bibitem{Zhou:2021bqs}
S.~Zhou,
``The smallest neutrino mass revisited,''
JHEP \textbf{11} (2021), 101
[arXiv:2104.09050 [hep-ph]].

\bibitem{Hyper-KamiokandeProto-:2015xww}
K.~Abe \textit{et al.} [Hyper-Kamiokande Proto-],
``Physics potential of a long-baseline neutrino oscillation experiment using a J-PARC
neutrino beam and Hyper-Kamiokande,''
PTEP \textbf{2015} (2015), 053C02
[arXiv:1502.05199 [hep-ex]].

\bibitem{DUNE:2015lol}
R.~Acciarri \textit{et al.} [DUNE],
``Long-Baseline Neutrino Facility (LBNF) and Deep Underground Neutrino Experiment (DUNE):
Conceptual Design Report, Volume 2: The Physics Program for DUNE at LBNF,''
[arXiv:1512.06148 [physics.ins-det]].

\bibitem{Fukuyama:2020swd}
T.~Fukuyama and Y.~Mimura,
``$\mu$-$\tau$ symmetry breaking and CP violation in the neutrino mass matrix,''
Phys. Rev. D \textbf{102} (2020) no.1, 016002
[arXiv:2001.11185 [hep-ph]].

\bibitem{Endoh:2002wm}
T.~Endoh, S.~Kaneko, S.~K.~Kang, T.~Morozumi and M.~Tanimoto,
``CP violation in neutrino oscillation and leptogenesis,''
Phys. Rev. Lett. \textbf{89} (2002), 231601
[arXiv:hep-ph/0209020 [hep-ph]].

\bibitem{Vafa:2005ui}
C.~Vafa,
``The String landscape and the swampland,''
[arXiv:hep-th/0509212 [hep-th]].

\bibitem{Palti:2019pca}
E.~Palti,
``The Swampland: Introduction and Review,''
Fortsch. Phys. \textbf{67} (2019) no.6, 1900037
[arXiv:1903.06239 [hep-th]].

\bibitem{Xing:2021SUSY}
Z.~z.~Xing,
``Neutrino masses and Yukawa interactions,''
Plenary talk given at the 28th International Conference on Supersymmetry and
Unification of Fundamental Interactions (SUSY 2021),
August 2021, Beijing.

\end{thebibliography}
\end{document}